\documentclass[aps,10pt,onecolumn,notitlepage,nofootinbib,longbibliography]{revtex4-2}
\usepackage{amsmath}
\usepackage{amssymb}
\usepackage{graphicx}
\usepackage{epsdice}
\usepackage[T1]{fontenc}
\usepackage{hyperref}
\usepackage{url}
\makeatletter
\renewcommand{\paragraph}{%
  \@startsection{paragraph}{4}{\parindent}{\z@}{-1em}%
  {\normalfont\normalsize\bfseries}%
}
\makeatother

\begin{document}

\title{Foundations of entropy in complex systems}
\author{Jan Korbel}
\email{korbel@csh.ac.at}
\thanks{ORCID: \url{https://orcid.org/0000-0002-5371-5320}}
\affiliation{Complexity Science Hub, Metternichgasse 8, 1030 Vienna, Austria}
\date{\today}

\begin{abstract}
This chapter reviews the foundations of entropy and their extensions to complex systems. We first discuss the relation between Boltzmann's formula, multiplicity, coarse-graining, and Shannon entropy, before introducing generalized entropies such as R\'enyi, Tsallis, and Burg entropy. We then examine Maxwell--Boltzmann, Bose--Einstein, and Fermi--Dirac statistics, structure-forming systems, sample-space reducing processes, P\'olya urns, and nonlinear dynamics. Axiomatic approaches are presented through the Shannon--Khinchin axioms, Tempesta group-composability, Hanel--Thurner asymptotic scaling, Shore--Johnson consistency axioms, and Lieb--Yngvason axioms. Finally, we discuss calibration invariance, Hanel--Thurner--Gell-Mann duality between linear and escort averages, and Kolmogorov--Nagumo averages, showing how the same distribution can arise from different entropies, constraints, or dynamics. These results emphasize that the choice of entropy should be guided by the structure and physical properties of the system.
\end{abstract}

\maketitle

\vspace{0.1cm}
\begin{center}    
\includegraphics[width=0.5\linewidth]{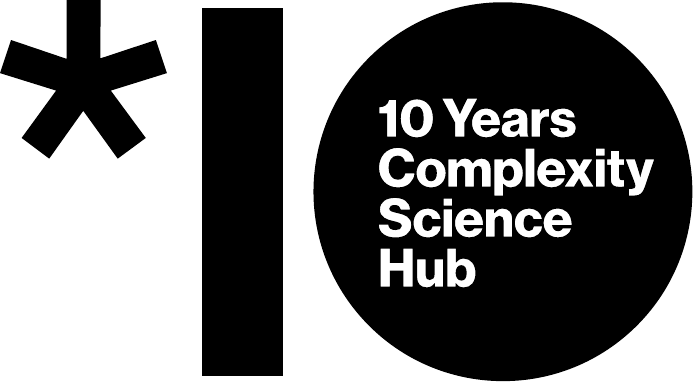}

\emph{This chapter is dedicated to the tenth anniversary of the Complexity Science Hub and, in particular, to the July 2026 Research Spotlight Month on the foundations of complex systems.}
\end{center}

\tableofcontents

\section{Introduction}
The main aim of physics is to predict how physical systems behave and evolve. Such predictions are commonly formulated in terms of a set of variables known as degrees of freedom. Examples include the position \(x\), the velocity \(v\), or the angular momentum \(L\). The dynamics of each degree of freedom is typically governed by a differential equation coupled to those describing the other degrees of freedom. For systems involving only a few variables, these equations can often be solved exactly.

A direct microscopic description, however, becomes impractical for macroscopic systems. A single mole of matter contains approximately \(6.022\times10^{23}\) constituent particles, making it impossible to specify and solve the equations of motion for every microscopic degree of freedom. The description of such systems therefore requires coarse-graining: microscopic configurations are merged into macroscopic states characterized by a comparatively small number of observable quantities.

The concept of entropy emerged from the thermodynamic description of macroscopic systems and acquired a statistical interpretation through the work of Boltzmann and Gibbs. In particular, Gibbs developed a systematic statistical-mechanical framework based on ensembles of possible microscopic states \cite{Gibbs1902statistical}. For a system with \(N\) discrete states, and probability distribution \(\boldsymbol{p}=(p_1,\ldots,p_N)\), the Shannon entropy is
\begin{equation}
H(\boldsymbol{p}) = -\sum_{i=1}^{N} p_i \ln p_i .
\end{equation}
Its information-theoretic interpretation quantifies the uncertainty associated with the distribution. It is connected with the physical Boltzmann--Gibbs entropy via multiplication by the Boltzmann constant \(k_{\mathrm B}=1.381 \times 10^{-23} J/K\), i.e.,
\begin{equation}
S(\boldsymbol{p}) = k_{\mathrm B} H(\boldsymbol{p}) .
\end{equation}
In the remainder of the chapter, we use units in which \(k_{\mathrm B}=1\), unless stated otherwise.
We use \(H\) specifically for Shannon entropy, \(S\) for a general entropy functional, and \(\mathcal{S}_{X}\) for the entropy per particle or time step associated with a particular framework \(X\).

The classical statistical-mechanical framework is particularly successful for systems composed of many identical and weakly interacting particles. Many systems encountered in nature and society, however, are heterogeneous, strongly interacting, history-dependent, or driven far from equilibrium. Their collective behavior may produce emergent states that cannot be understood solely from the properties of their individual components. Such systems form the subject of complexity science \cite{Thurner2018complex}.

This raises the central question addressed in this chapter: whether, and under what conditions, the classical concept of entropy can be extended to describe heterogeneous, strongly interacting systems with emergent collective behavior. We approach this question by first illustrating the broad applications and prominent generalizations of entropy. We then examine how entropy emerges from the multiplicity of accessible states, compare this microscopic perspective with axiomatic approaches, and finally investigate dualities and invariances of MaxEnt distributions in complex systems.

\subsection{Applications of entropy in various disciplines}
Although entropy was originally established in thermodynamics, it has developed into a broadly applicable concept for describing uncertainty, diversity, irreversibility, and the production of information. Its precise interpretation depends on how the system is described, but closely related entropy measures now appear across the natural, computational, life, and social sciences.

In physics, entropy was originally defined as a thermodynamic state function associated with equilibrium and the direction of spontaneous processes \cite{clausius1879mechanical}. Statistical mechanics connects this macroscopic quantity to the number and probabilities of microscopic configurations \cite{Gibbs1902statistical}. In gravitation and cosmology, black holes and cosmological event horizons possess entropies proportional to their surface areas, revealing a fundamental relation between thermodynamics, gravity, and information \cite{Bekenstein1873prd,gibbons1977prd}.

Entropy is equally central to chemistry. Together with energy, it determines the thermodynamic potentials that govern chemical equilibrium, phase coexistence, mixing, and the spontaneous direction of chemical transformations \cite{Gibbs1878equilibrium}. In the theory of dynamical systems, quantities such as Kolmogorov--Sinai entropy and topological entropy quantify the production of information and the growth of distinguishable trajectories, providing measures of dynamical complexity and chaos \cite{Downarowicz2011entropy}. Quantum theory replaces classical probability distributions with density operators; the corresponding von Neumann entropy quantifies uncertainty and mixedness and plays a central role in the study of entanglement and quantum information \cite{Nielsen2010quantum}.

In information theory, Shannon entropy measures the uncertainty of a source and establishes fundamental limits on data compression and reliable communication \cite{Shannon1948bell,Cover2006information}. These ideas directly enter computer science through entropy coding and related compression methods. Entropy-based quantities also appear throughout statistical inference and machine learning; in particular, cross-entropy is widely used as an objective function for training classification models \cite{Goodfellow2016deep}.

Applications in the life sciences use entropy to characterize organization, diversity, and information processing. Early discussions emphasized the relation between physical entropy and biological order \cite{schrodinger1944what}. More quantitative applications include measuring the information content of nucleotide binding sites \cite{Schneider1986information} and the entropy and information carried by neural spike trains \cite{Strong1998entropy}. In ecology, Shannon entropy provides a diversity index that accounts for both the number of species and their relative abundances \cite{magurran2013ecological,jost2006entropy}. In medicine and physiology, entropy is used to quantify irregularity in physiological time series, including heart-rate dynamics \cite{Richman2000physiological}.

Entropy has also become useful in the study of human and social systems. In linguistics, it quantifies the predictability and redundancy of natural language \cite{Shannon1951bell}. In economics, the entropy-derived Theil index measures inequality in income and wealth distributions \cite{theil1967economics}. In finance, entropy can serve as an alternative measure of portfolio diversification and market risk \cite{Philippatos1972entropy}. These applications show that entropy is not restricted to a specific physical system and can quantify the distribution and accessibility of possible states in many different real-world systems.

\subsection{Generalized entropies}

The Shannon--Boltzmann--Gibbs entropy is appropriate for a broad range of systems, but its derivation as an extensive thermodynamic entropy relies on assumptions about the composition and multiplicity of accessible states. Complex systems can violate these assumptions due to strong correlations, long-range interactions, history dependence, or constraints on the growth of their configuration spaces. Generalized entropies modify how probabilities contribute to uncertainty and provide alternative composition rules that can better reflect such structures. The parameters appearing in these entropies often control the relative weight assigned to rare and common events. In this section, we briefly review some of the most prominent examples of generalized entropies:

\paragraph{Tsallis-Havrda-Charv\'{a}t entropy.} One of the most widely used examples is the Tsallis (also called Tsallis--Havrda--Charvát) entropy,
\begin{equation}
S_q(\boldsymbol{p})=\frac{\sum_i p_i^q-1}{1-q}.
\end{equation}
It was introduced in information theory \cite{havrda1967kybernetika,daroczy1970infcont} and later developed as the basis of nonextensive statistical mechanics \cite{Tsallis1988jstatphys,Tsallis2009nonextensive}. For \(q>0\), it is concave and recovers the Shannon entropy in the limit \(q\to1\). Values \(0<q<1\) give greater
relative weight to rare events, whereas \(q>1\) emphasizes events with high probability. For two statistically independent systems \(A\) and \(B\), it satisfies the pseudo-additive composition rule
\begin{equation}
S_q(A,B)=S_q(A)+S_q(B)+(1-q)S_q(A)S_q(B).
\end{equation}
Maximizing \(S_q\) under suitable constraints gives rise to \(q\)-exponential distributions, which can exhibit power-law tails. The entropy has consequently been applied to systems involving long-range interactions, anomalous diffusion, metastable states, and heavy-tailed distributions.

\paragraph{R\'{e}nyi entropy.} Another fundamental family is the R\'enyi entropy,
\begin{equation}
R_q(\boldsymbol{p})=\frac{1}{1-q}\ln\left(\sum_i p_i^q\right),
\end{equation}
which was introduced as a generalized measure of information \cite{renyi1961proceedings}. It is commonly defined for \(q>0\), with its Shannon limit obtained as \(q\to1\). For fixed \(q\), it is a monotonic function of the Tsallis entropy and is Schur-concave, ensuring that it increases under the mixing of a probability distribution. In contrast to the Tsallis entropy, the R\'enyi entropy is additive for statistically independent systems. Varying \(q\) changes the relative sensitivity to common and rare events, making the family particularly useful in multifractal analysis \cite{jizba2004annphys}, information theory, and quantum information \cite{Nielsen2010quantum}.

\paragraph{Kaniadakis entropy.} The Kaniadakis entropy is based on the deformed \(\kappa\)-logarithm and is defined as
\begin{equation}
K_\kappa(\boldsymbol{p})
=-\sum_i p_i\ln_\kappa p_i
=\sum_i\frac{p_i^{1-\kappa}-p_i^{1+\kappa}}{2\kappa}.
\end{equation}
For the conventional range \(-1<\kappa<1\), it is symmetric under \(\kappa\to-\kappa\) and approaches the Shannon entropy as \(\kappa\to0\). The associated \(\kappa\)-exponential distributions have power-law tails, and the deformation arises naturally in a statistical-mechanical formulation compatible with special relativity \cite{kaniadakis2001physa,Kaniadakis2002physa,Kaniadakis2002pre}.

\paragraph{Sharma-Mittal entropy.} The two-parameter Sharma--Mittal entropy,
\begin{equation}
S_{r,q}(\boldsymbol{p})
=\frac{\left(\sum_i p_i^q\right)^{\frac{1-r}{1-q}}-1}{1-r},
\end{equation}
provides a broader family that interpolates between several generalized entropies \cite{Sharma1975metrika,Mittal1975metrika}. It is commonly considered for \(q>0\), while the admissible range of \(r\) depends on the desired mathematical and thermodynamic properties. It reduces to the R\'enyi entropy in the limit \(r\to1\) and to the Tsallis entropy when \(r\to q\). Its two parameters separately control the weighting of probabilities and the entropy's composition properties.

\paragraph{Burg entropy.} Finally, an example with a different structure is the Burg entropy,
\begin{equation}
S_{\mathrm{BU}}(\boldsymbol{p})=\sum_i\ln p_i,
\end{equation}
up to a convention-dependent normalization. It strongly penalizes probabilities approaching zero and diverges when any \(p_i\to0\), distinguishing it from entropy forms that are unchanged by the addition of an impossible event. It has important applications in maximum-entropy spectral estimation, where it is closely related to likelihood-based inference \cite{Burg1972geo}.

These examples represent only a small subset of the generalized entropies proposed in information theory and statistical physics \cite{Amigo2018entropy,Ilic2021epl,Beck2008contphys}. Such entropies can be constructed from deformed logarithms, modified composition laws, or axiomatic requirements, while generalized MaxEnt principles can also arise from curved statistical manifolds \cite{PhysRevResearch.3.033216}. In this chapter, we focus on their relation to the multiplicity of accessible configurations: changes in the combinatorial structure of a system can lead to entropy forms different from the Shannon--Boltzmann--Gibbs entropy \cite{hanel2014pnas}.

\section{Boltzmann's entropy}

We now return to the statistical-mechanical origin of entropy and develop the multiplicity-based perspective that will be used throughout this chapter. The starting point is Boltzmann's formula,
\begin{equation}
S = k_{\mathrm B}\ln W,
\end{equation}
where \(W\) denotes the multiplicity, that is, the number of microscopic configurations compatible with a given coarse-grained description of the system. Once the relevant configurations and their multiplicities are identified, Boltzmann's formula determines the entropy and the most probable macroscopic behavior.

For systems whose configurations are generated by independent sampling, the multiplicity is multinomial and Boltzmann's formula leads to the Shannon--Boltzmann--Gibbs entropy. Complex systems, however, may possess correlations, path dependence, or constraints that produce non-multinomial multiplicities. Applying the same Boltzmann relation to these systems naturally gives rise to non-Shannonian entropy forms. We therefore begin by distinguishing different levels of description and then examine how multiplicity determines entropy in the thermodynamic limit.

\subsection{Microstates, Mesostates and Macrostates}

We begin by distinguishing three levels at which a many-body system can be described. Consider a system composed of \(n\) distinguishable constituents, each of which can occupy one of \(N\) states from the set
\begin{equation}
\mathbb{S}=\{s_i\}_{i=1}^{N}.
\end{equation}
Depending on the system, these states may represent spin orientations, spatial cells, energy levels, chemical species, or other microscopic alternatives. A \emph{microstate} specifies the state of every constituent and can therefore be represented by an ordered sequence
\begin{equation}
\boldsymbol{x}=(x_1,\ldots,x_n), \qquad x_j\in\mathbb{S}.
\end{equation}

At a coarser level, let \(n_i \geq 0\) denote the number of constituents occupying state \(s_i\). The occupation numbers satisfy
\begin{equation}
\sum_{i=1}^{N}n_i=n,
\end{equation}
and their frequency histogram
\begin{equation}
\boldsymbol{n}=(n_1,\ldots,n_N)
\end{equation}
defines a \emph{mesostate}. A mesostate retains the occupation numbers but discards information about which individual constituent occupies each state. Finally, a \emph{macrostate} is specified by one or more macroscopic observables obtained from the microstate or mesostate, such as the total energy, magnetization, or, more generally, a mean value of a measured quantity. Each successive step is produced by coarse-graining and generally corresponds to many configurations at the finer level.

As a simple example, consider five throws of a six-sided die, for which \(\mathbb{S}=\{\epsdice{1},\epsdice{2},\epsdice{3},\epsdice{4},\epsdice{5},\epsdice{6}\}\). Suppose the observed sequence is
\begin{equation}
\scalebox{1.4}{\(\epsdice{4}\)}
\quad
\scalebox{1.4}{\(\epsdice{2}\)}
\quad
\scalebox{1.4}{\(\epsdice{6}\)}
\quad
\scalebox{1.4}{\(\epsdice{2}\)}
\quad
\scalebox{1.4}{\(\epsdice{5}\)}
\end{equation}
or equivalently \(\boldsymbol{x}=(4,2,6,2,5)\). This ordered sequence is a microstate. Its histogram is
\begin{equation}
\boldsymbol{n}=(0,2,0,1,1,1),
\end{equation}
which defines the corresponding mesostate: the outcome \(2\) occurs twice, the outcomes \(4\), \(5\), and \(6\) occur once, and the remaining outcomes do not occur. If only the sample mean is observed, the system is described by the macrostate
\begin{equation}
\overline{x}=\frac{1}{5}\sum_{j=1}^{5}x_j
=\frac{4+2+6+2+5}{5}=3.8.
\end{equation}

This example also illustrates the reduction in the number of distinguishable states under coarse-graining. There are
\begin{equation}
6^5=7776
\end{equation}
possible ordered microstates. The number of possible histograms of five throws among six outcomes is given by
\begin{equation}
\binom{N+n-1}{n}=\binom{6+5-1}{5}=252.
\end{equation}
The possible sums of five dice outcomes range from \(5\) to \(30\), giving \(26\) distinct values of the mean and therefore \(26\) macrostates under this particular macroscopic description. 

Indeed, many different microstates can correspond to the same mesostate. The number of microstates associated with a given mesostate \(\boldsymbol{n}\) is called its \emph{multiplicity} and is denoted by \(W(\boldsymbol{n})\). Consider first the dice mesostate \(\boldsymbol{n}=(0,2,0,1,1,1)\). There are \(5!\) permutations of the five throws. However, exchanging the two occurrences of \(\epsdice{2}\) does not change the resulting microstate, leading to an overcounting by a factor of \(2!\). Its multiplicity is therefore
\begin{equation}
W(0,2,0,1,1,1)
=\frac{5!}{0!\,2!\,0!\,1!\,1!\,1!}
=\frac{5!}{2!}
=60.
\end{equation}
Thus, exactly \(60\) of the \(6^5=7776\) possible ordered sequences share this histogram.

The same argument applies in general. We first consider all \(n!\) permutations of the \(n\) constituents. Whenever \(n_i\) constituents occupy the same state \(s_i\), their \(n_i!\) internal permutations do not produce distinct microstates and must therefore be divided out. The multiplicity of a mesostate is consequently given by the multinomial coefficient
\begin{equation}
W(\boldsymbol{n})
=\binom{n}{n_1,\ldots,n_N}
=\frac{n!}{\prod_{i=1}^{N}n_i!}.
\end{equation}
The multiplicity quantifies how much microscopic information is discarded when a microstate is coarse-grained into a mesostate and provides the combinatorial foundation of Boltzmann's entropy.

We can now derive the Shannon--Boltzmann--Gibbs entropy directly from Boltzmann's formula. Setting \(k_{\mathrm B}=1\) for the moment and substituting the multinomial multiplicity gives
\begin{equation}
S(\boldsymbol{n})=\ln W(\boldsymbol{n})
=\ln n!-\sum_{i=1}^{N}\ln n_i!.
\end{equation}
For large occupation numbers, Stirling's approximation, \(\ln x!\simeq x\ln x-x\), yields
\begin{equation}
S(\boldsymbol{n})
\simeq n\ln n-n-\sum_{i=1}^{N}\left(n_i\ln n_i-n_i\right)
=n\ln n-\sum_{i=1}^{N}n_i\ln n_i,
\end{equation}
where we used \(\sum_i n_i=n\). Introducing the relative-frequency vector \(\boldsymbol{p}=(p_1,\ldots,p_N)\), with components
\begin{equation}
p_i=\frac{n_i}{n},
\end{equation}
and substituting \(n_i=np_i\), we obtain
\begin{equation}
S(\boldsymbol{n})
\simeq -n\sum_{i=1}^{N}p_i\ln p_i.
\end{equation}
The entropy per constituent is therefore
\begin{equation}
\frac{S(\boldsymbol{n})}{n}
\simeq -\sum_{i=1}^{N}p_i\ln p_i
=H(\boldsymbol{p}).
\end{equation}
Here and throughout, the boundary terms are defined by the convention \(0\cdot \ln 0=0\).
Restoring the Boltzmann constant gives the physical entropy per constituent as \(k_{\mathrm B}H(\boldsymbol{p})\). Thus, the Shannon--Boltzmann--Gibbs entropy emerges from Boltzmann's formula when the multiplicity is multinomial, and the system is considered in the large \(n\) limit.

\subsection{Multiplicity in the thermodynamic limit}

The preceding derivation introduced the relative frequencies \(p_i=n_i/n\). For a finite system, these frequencies fluctuate from one realization to another and many different mesostates can be compatible with the same macroscopic observation. When compatible microstates are assigned equal probability, the distribution of mesostates becomes sharply concentrated around those with maximal multiplicity in the thermodynamic limit \(n\to\infty\). Large-deviation theory (LDT) provides the general mathematical framework for this exponential concentration and its connection to equilibrium statistical mechanics \cite{Ellis1985large}. For nonuniform independent sampling, the probability of a mesostate also contains the sampling weights, and the corresponding concentration is governed by relative entropy rather than multiplicity alone. This concentration is the reason why a macroscopic system can be described by a small number of reproducible observables despite its enormous number of microscopic degrees of freedom.

To illustrate this point, let us return to the example of five six-sided dice and suppose that only the mean value
\begin{equation}
\overline{x}=3.8
\end{equation}
is observed. Equivalently, the sum of the five outcomes is \(19\). Many mesostates satisfy this constraint. For example, the ordered representations
\begin{equation}
(6,5,4,3,1),\qquad
(6,5,4,2,2),\qquad
(4,4,4,4,3)
\end{equation}
all have the same sum and therefore correspond to the same macrostate. Their multiplicities, however, are very different:
\begin{eqnarray}
W(6,5,4,3,1)&=&5!=120,\\
W(6,5,4,2,2)&=&\frac{5!}{2!}=60,\\
W(4,4,4,4,3)&=&\frac{5!}{4!}=5.
\end{eqnarray}
Here, the tuples list the dice outcomes rather than the occupation-number histograms; permutations of a tuple belong to the same mesostate. If all microstates compatible with the macroscopic constraint are assigned equal probability, the probability of observing a particular mesostate is proportional to its multiplicity,
\begin{equation}
P(\boldsymbol{n}\mid \overline{x})
=\frac{W(\boldsymbol{n})}
{\sum_{\boldsymbol{n}'\,:\,\overline{x}(\boldsymbol{n}')=\overline{x}}
W(\boldsymbol{n}')}.
\end{equation}
Consequently, the first mesostate in the example is \(24\) times more probable than the last one, even though both produce the same macroscopic mean.

The difference becomes much more pronounced as the system is enlarged. Suppose a mesostate with occupation numbers \(\boldsymbol{n}=(n_1,\ldots,n_N)\) is rescaled by a factor \(\lambda\), so that the occupation numbers become \(\lambda\boldsymbol{n}\) while the relative frequencies \(p_i=n_i/n\) remain unchanged. Its multiplicity is
\begin{equation}
W(\lambda\boldsymbol{n})
=\frac{(\lambda n)!}{\prod_i(\lambda n_i)!}.
\end{equation}
Using Stirling's approximation gives the leading asymptotic behavior
\begin{equation}
\ln W(\lambda\boldsymbol{n})
\simeq \lambda n H(\boldsymbol{p}),
\end{equation}
or equivalently, using the LDT notation \cite{Ellis1985large},
\begin{equation}
W(\lambda\boldsymbol{n})
\asymp \exp\left[\lambda n H(\boldsymbol{p})\right].
\end{equation}
Thus, a difference in entropy between two mesostates becomes an exponential difference in their multiplicities. If \(\boldsymbol{p}^\star\) denotes the distribution with maximal entropy among all distributions satisfying the macroscopic constraints, then for any other admissible distribution \(\boldsymbol{p}\),
\begin{equation}
\frac{W(\lambda n\boldsymbol{p})}{W(\lambda n\boldsymbol{p}^\star)}
\asymp
\exp\left[-\lambda n\left(H(\boldsymbol{p}^\star)-H(\boldsymbol{p})\right)\right].
\end{equation}
Provided \(H(\boldsymbol{p})<H(\boldsymbol{p}^\star)\), this ratio vanishes exponentially as \(\lambda n\to\infty\). Mesostates away from the maximum therefore become negligible compared with the most probable one.
If several admissible distributions attain the same maximum, the concentration occurs around the set of maximizers rather than a single distribution.

This concentration also gives a statistical interpretation to the probabilities \(p_i\). For independent sampling, define the empirical frequencies by
\begin{equation}
\widehat{p}_i(n):=\frac{n_i(n)}{n}.
\end{equation}
The law of large numbers implies \(\widehat{p}_i(n)\to p_i\), so the empirical frequencies converge to the underlying sampling probabilities in the thermodynamic limit. For independent sampling with finite variance, typical fluctuations of the occupation numbers grow as \(\sqrt{n}\), and hence their relative size decays as \(1/\sqrt{n}\). Macroscopic observables therefore converge to sharply defined ensemble values. Determining the equilibrium macrostate can consequently be reduced to finding the mesostate of maximum multiplicity, or equivalently, maximizing its entropy subject to the relevant macroscopic constraints.

\subsection{MaxEnt distribution}

Finding the state with maximal multiplicity leads naturally to the principle of maximum entropy and MaxEnt distribution. Although maximizing multiplicity and applying MaxEnt as an inference principle are conceptually distinct procedures, for multinomial multiplicities with equally weighted microstates, they yield the same optimization problem \cite{Ellis1985large,PhysRev.106.620}: among all probability distributions consistent with the available information, one chooses the distribution with maximal Shannon entropy,
\begin{equation}
H(\boldsymbol{p})=-\sum_{i=1}^{N}p_i\ln p_i.
\end{equation}
Suppose that the states have energies \(\epsilon_i\) and that the available information consists of normalization and a fixed mean energy,
\begin{equation}
\sum_{i=1}^{N}p_i=1,\qquad p_i\geq0,
\qquad
\sum_{i=1}^{N}p_i\epsilon_i=U.
\end{equation}
We introduce Lagrange multipliers \(\alpha\) and \(\beta\) and define
\begin{equation}
\mathcal{L}(\boldsymbol{p})
=H(\boldsymbol{p})
-\alpha\left(\sum_{i=1}^{N}p_i-1\right)
-\beta\left(\sum_{i=1}^{N}p_i\epsilon_i-U\right).
\end{equation}
Stationarity with respect to each probability gives
\begin{equation}
\frac{\partial\mathcal{L}}{\partial p_i}
=-\ln p_i-1-\alpha-\beta\epsilon_i=0.
\end{equation}
Since the Shannon entropy is strictly concave on the probability simplex and the constraints are linear, this stationary point is the unique maximum. Solving for \(p_i\) and imposing normalization yields
\begin{equation}
p_i^\star=\frac{\exp(-\beta\epsilon_i)}{Z(\beta)},
\qquad
Z(\beta)=\sum_{i=1}^{N}\exp(-\beta\epsilon_i).
\end{equation}
Thus, maximizing Shannon entropy under a mean-energy constraint produces the canonical distribution of equilibrium statistical mechanics \cite{Gibbs1902statistical,PhysRev.106.620}. The normalization multiplier is absorbed into the partition function, while \(\beta\) is determined by the prescribed mean energy and is identified in equilibrium with \(1/(k_{\mathrm B}T)\).

More generally, if the available information is expressed through expectation values
\begin{equation}
\sum_i p_i I_{i,k}=\overline{I}_k,
\end{equation}
then maximizing Shannon entropy gives an exponential-family distribution,
\begin{equation}
p_i^\star
=\frac{1}{Z(\boldsymbol{\lambda})}
\exp\left(-\sum_k\lambda_k I_{i,k}\right),
\end{equation}
with
\begin{equation}
Z(\boldsymbol{\lambda})
=\sum_i\exp\left(-\sum_k\lambda_k I_{i,k}\right).
\end{equation}
The MaxEnt principle provides a general method for constructing the least informative distribution compatible with specified constraints. For multinomial multiplicities with equally weighted microstates, this construction recovers the canonical ensemble without requiring an explicit model of a heat bath.

\section{Entropies for systems beyond multinomial multiplicity}

We now turn to examples that go beyond the standard correspondence between multinomial multiplicity and Shannon entropy. This can occur through modified microscopic counting rules, through history-dependent trajectory multiplicities or probabilities, or by deforming the combinatorial operations themselves. In each case, we examine how the underlying statistical structure determines the corresponding entropy and most probable distribution.

\subsection{Maxwell--Boltzmann, Bose--Einstein, and Fermi--Dirac statistics}

Maxwell--Boltzmann (MB), Bose--Einstein (BE), and Fermi--Dirac (FD) distributions are often introduced through the grand-canonical ensemble. An alternative route starts directly from the combinatorial rules that distinguish the three statistics. Once the correct multiplicity has been identified, Boltzmann's formula determines the corresponding entropy, and its constrained maximization yields the familiar equilibrium distribution. In this comparison, MB statistics provides the multinomial reference case, whereas the indistinguishability and occupancy restrictions of BE and FD statistics produce different multiplicities. This multiplicity-first approach also admits exact finite-size entropy and most-probable distributions without invoking Stirling's approximation \cite{niven2005physletta,niven2009epjb}. Here, we focus on the thermodynamic-limit forms in order to make the connection between multiplicity, entropy, and MaxEnt explicit.

Consider \(n\) particles distributed among \(N\) distinguishable energy levels with energies \(\epsilon_i\). Let \(n_i\) be the number of particles occupying level \(i\), and let \(g_i\) denote the number of distinguishable sublevels associated with that level. We introduce
\begin{equation}
p_i=\frac{n_i}{n},
\qquad
\eta_i=\frac{g_i}{n},
\end{equation}
with the constraints
\begin{equation}
\sum_{i=1}^{N}p_i=1,
\qquad
\sum_{i=1}^{N}p_i\epsilon_i=U.
\end{equation}
The three statistics differ in whether particles are distinguishable and in how many particles may occupy each sublevel.
For the thermodynamic-limit expressions below, \(n_i,g_i\to\infty\) while \(p_i=n_i/n\) and \(\eta_i=g_i/n\) remain fixed.

\paragraph{Maxwell--Boltzmann statistics.}
In MB statistics, the particles are distinguishable and any number of particles may occupy a sublevel. For a fixed occupation vector \(\boldsymbol{n}\), the particles can first be assigned to the energy levels in \(n!/\prod_i n_i!\) ways. Each of the \(n_i\) particles in level \(i\) can then occupy any of its \(g_i\) sublevels. The multiplicity is therefore
\begin{equation}
W_{\mathrm{MB}}(\boldsymbol{n})
=n!\prod_{i=1}^{N}\frac{g_i^{n_i}}{n_i!}.
\end{equation}
Using Boltzmann's relation and Stirling's approximation, the entropy per particle becomes, up to terms independent of \(\boldsymbol{p}\),
\begin{equation}
\mathcal{S}_{\mathrm{MB}}(\boldsymbol{p})
=-\sum_{i=1}^{N}p_i\ln\frac{p_i}{\eta_i}.
\end{equation}
Its stationarity condition is
\begin{equation}
\frac{\partial\mathcal{S}_{\mathrm{MB}}}{\partial p_i}
=-\ln\frac{p_i}{\eta_i}-1
=\alpha+\beta\epsilon_i,
\end{equation}
where \(\alpha\) is the normalization multiplier. Solving this equation and imposing normalization gives
\begin{equation}
p_i^\star
=\frac{\eta_i}{Z_{\mathrm{MB}}}
\exp(-\beta\epsilon_i),
\qquad
Z_{\mathrm{MB}}=\sum_i \eta_i\exp(-\beta\epsilon_i).
\end{equation}
Thus, the degeneracy factor appears directly from the multiplicity rather than being added to the canonical distribution afterward.

\paragraph{Bose--Einstein statistics.}
Bosons are indistinguishable, and any number of particles may occupy the same sublevel. The number of ways to distribute \(n_i\) indistinguishable particles among \(g_i\) distinguishable sublevels is given by the stars-and-bars coefficient. Consequently,
\begin{equation}
W_{\mathrm{BE}}(\boldsymbol{n})
=\prod_{i=1}^{N}
\binom{n_i+g_i-1}{n_i}.
\end{equation}
In the thermodynamic limit, Boltzmann's relation gives the entropy per particle
\begin{equation}
\mathcal{S}_{\mathrm{BE}}(\boldsymbol{p})
=\sum_{i=1}^{N}
\left[
(\eta_i+p_i)\ln(\eta_i+p_i)
-\eta_i\ln\eta_i
-p_i\ln p_i
\right].
\end{equation}
The stationarity condition for constrained maximization is
\begin{equation}
\frac{\partial\mathcal{S}_{\mathrm{BE}}}{\partial p_i}
=\ln\frac{\eta_i+p_i}{p_i}
=\beta(\epsilon_i-\mu),
\end{equation}
which yields
\begin{equation}
p_i^\star
=\frac{\eta_i}
{\exp[\beta(\epsilon_i-\mu)]-1},
\end{equation}
where the chemical potential \(\mu\) is determined by normalization. The minus sign in the denominator follows directly from the unrestricted occupation of each bosonic sublevel.

\paragraph{Fermi--Dirac statistics.}
Fermions are indistinguishable and obey the Pauli exclusion principle, so each sublevel can contain at most one particle. The occupation numbers must therefore satisfy \(n_i\leq g_i\), and the number of ways to select the occupied sublevels is
\begin{equation}
W_{\mathrm{FD}}(\boldsymbol{n})
=\prod_{i=1}^{N}
\binom{g_i}{n_i}.
\end{equation}
The corresponding thermodynamic-limit entropy per particle is
\begin{equation}
\mathcal{S}_{\mathrm{FD}}(\boldsymbol{p})
=\sum_{i=1}^{N}
\left[
-(\eta_i-p_i)\ln(\eta_i-p_i)
+\eta_i\ln\eta_i
-p_i\ln p_i
\right],
\end{equation}
defined for \(0\leq p_i\leq\eta_i\). In this case, the stationarity condition becomes
\begin{equation}
\frac{\partial\mathcal{S}_{\mathrm{FD}}}{\partial p_i}
=\ln\frac{\eta_i-p_i}{p_i}
=\beta(\epsilon_i-\mu),
\end{equation}
and its constrained maximum is
\begin{equation}
p_i^\star
=\frac{\eta_i}
{\exp[\beta(\epsilon_i-\mu)]+1}.
\end{equation}
The plus sign in the denominator is the statistical consequence of exclusion.

The MB, BE, and FD distributions can therefore all be obtained from the same principle: identify the correct combinatorial multiplicity and maximize its logarithm under the relevant constraints. Their different equilibrium forms arise not from changing the extremization procedure, but from changing the microscopic counting rules. In the dilute limit, where \(p_i\ll\eta_i\), both the BE and FD distributions approach the MB form. The exact expressions discussed by Niven retain the factorials in the multiplicities and replace the Stirling-based stationarity equations by finite-size relations involving gamma and digamma functions \cite{niven2005physletta,niven2009epjb}.

\paragraph{Gentile statistics.}
Gentile statistics interpolates between FD and BE statistics by imposing a finite maximum occupation number \(r\) on each sublevel \cite{niven2009epjb,Dai2004gentile}. For a level \(i\), the multiplicity is the coefficient of \(x^{n_i}\) in \((1+x+\cdots+x^r)^{g_i}\). By the inclusion--exclusion principle, it can be written as
\begin{equation}
W_{\mathrm{G},i}(n_i,g_i,r)
=
\sum_{k=0}^{\left\lfloor n_i/(r+1)\right\rfloor}
(-1)^k
\binom{g_i}{k}
\binom{n_i-k(r+1)+g_i-1}{g_i-1},
\end{equation}
and the total multiplicity is \(W_{\mathrm G}(\boldsymbol{n})=\prod_iW_{\mathrm{G},i}(n_i,g_i,r)\). For \(r=1\), the multiplicity reduces to the FD coefficient \(\binom{g_i}{n_i}\). In the limit \(r\to\infty\), only the \(k=0\) term remains, yielding the BE coefficient \(\binom{n_i+g_i-1}{n_i}\).

Unlike the BE and FD multiplicities, the Gentile multiplicity is generally expressed as an alternating sum of binomial coefficients. Consequently, Boltzmann's formula involves the logarithm of a sum,
\begin{equation}
S_{\mathrm G}(\boldsymbol{n})
=
\sum_i\ln W_{\mathrm{G},i}(n_i,g_i,r),
\end{equation}
which cannot generally be decomposed into a simple trace-form entropy. Only in the limiting FD and BE cases does the sum reduce to a single binomial coefficient. Gentile statistics, therefore, illustrate that a well-defined multiplicity need not correspond to a simple closed-form entropy functional.

\subsection{Structure-forming systems}

Many systems consist of elementary constituents that can bind together and form emergent structures. Examples include atoms forming molecules, colloidal particles forming clusters \cite{C0SM00995D}, or individuals forming social groups \cite{Korbel2023prl}. Because a structure containing several particles acts as a new collective unit, the number of accessible configurations can grow faster than in systems composed only of independent particles \cite{Jensen2018jphysa}, i.e., faster than exponentially. These are called \emph{superexponential} systems. The corresponding multiplicity must account not only for the states occupied by the structures, but also for the different ways in which elementary constituents can be grouped. A thermodynamic description based on this counting was developed in Ref.~\cite{Korbel2021natcom}, building on a combinatorial idea already considered by Boltzmann \cite{boltzmann1884annalen}.

Consider a system of \(n\) distinguishable elementary particles that can form structures (which we call \emph{molecules}) of sizes \(j=1,\ldots,m\), where \(j=1\) denotes a free particle. A molecule of size \(j\) can occupy one of the states
\begin{equation}
\mathbb{S}^{(j)}
=\left\{s_1^{(j)},\ldots,s_{N_j}^{(j)}\right\}.
\end{equation}
A microstate must specify the state of every particle as well as the identities of the particles belonging to each structure. For example, a particle belonging to a two-particle molecule in state \(s_i^{(2)}\) can be represented by \(s_i^{(2)}(x_1,x_2)\), where \(x_1\) and \(x_2\) identify its two particles.

At the mesoscopic level, let \(n_i^{(j)}\) denote the number of molecules of size \(j\) in state \(s_i^{(j)}\). The mesostate is the occupation array
\begin{equation}
\boldsymbol{n}
=\left(n_i^{(j)}\right)_{i,j},
\end{equation}
satisfying conservation of particles
\begin{equation}
\sum_{j=1}^{m}\sum_{i=1}^{N_j}j n_i^{(j)}=n.
\end{equation}
This constraint differs from the usual normalization because a molecule of size \(j\) contains \(j\) elementary particles.

To calculate the multiplicity, we first permute all \(n\) distinguishable particles. We then divide by \(n_i^{(j)}!\) because molecules of the same size and state are indistinguishable at the mesoscopic level. We must also divide by \(j!\) for every molecule of size \(j\), since permutations of its constituent particles do not produce a new molecule. The resulting multiplicity is therefore
\begin{equation}
W(\boldsymbol{n})
=
\frac{n!}
{\displaystyle
\prod_{j=1}^{m}\prod_{i=1}^{N_j}
n_i^{(j)}!\,
(j!)^{n_i^{(j)}}}.
\end{equation}
The factors \((j!)^{n_i^{(j)}}\) encode the emergent molecule structure and distinguish this multiplicity from the multinomial coefficient.

We introduce the molecule ``probabilities''
\begin{equation}
\wp_i^{(j)}=\frac{n_i^{(j)}}{n}.
\end{equation}
These molecule ``probabilities''  \(\boldsymbol{\wp}
=\left(\wp_i^{(j)}\right)_{i,j}\) do not sum up to one in general, but satisfy 
\begin{equation}
\sum_{j=1}^{m}\sum_{i=1}^{N_j}j\wp_i^{(j)}=1.
\end{equation}
Applying Boltzmann's formula and Stirling's approximation to the multiplicity gives the entropy per elementary particle
\begin{eqnarray}
\mathcal{S}_{\mathrm{SF}}(\boldsymbol{\wp})
&=&
-\sum_{i,j}\wp_i^{(j)}
\left(\ln\wp_i^{(j)}-1\right)
\nonumber\\
&&
-\sum_{i,j}\wp_i^{(j)}
\ln\left(\frac{j!}{n^{j-1}}\right).
\end{eqnarray}
As is standard in MaxEnt calculations, terms independent of the distribution are omitted; in particular, the direct Stirling expansion contains an additional constant \(-1\). The entropy and the corresponding free energy are therefore stated up to this additive convention. The first term resembles Shannon entropy, but it is evaluated on molecule ``probabilities'' with a modified normalization. The second term accounts for the combinatorial cost of assembling \(j\) elementary particles into one molecule and for the rapidly growing number of possible groupings.

Let \(\epsilon_i^{(j)}\) be the energy of a structure of size \(j\) in state \(s_i^{(j)}\). The mean energy per elementary particle is defined as
\begin{equation}
\sum_{i,j}\epsilon_i^{(j)}\wp_i^{(j)}=U.
\end{equation}
Maximizing \(\mathcal{S}_{\mathrm{SF}}(\boldsymbol{\wp})\) under the particle-conservation and mean-energy constraints gives
\begin{equation}
\wp^\star{}_i^{(j)}
=
\frac{n^{j-1}}{j!}
\exp\left[-\alpha j-\beta\epsilon_i^{(j)}\right],
\end{equation}
where \(\alpha\) and \(\beta\) are determined by the constraints. The factor \(n^{j-1}/j!\) follows directly from the structure-forming multiplicity and modifies the ordinary exponential distribution by assigning a size-dependent statistical weight to each collective state.

Unlike in the ordinary canonical ensemble, normalization is not achieved by simply calculating a single partition function. Introducing the partial partition functions
\begin{equation}
\mathcal{Z}_j
=\frac{n^{j-1}}{j!}
\sum_i\exp\left(-\beta\epsilon_i^{(j)}\right)
\end{equation}
and \(z=\exp(-\alpha)\), the particle-conservation constraint becomes
\begin{equation}
\sum_{j=1}^{m}j\mathcal{Z}_jz^j=1.
\end{equation}
This is a polynomial equation of order \(m\) for the effective fugacity \(z\), which must generally be solved to normalize the MaxEnt distribution. The number of molecules per elementary particle is
\begin{equation}
\mathcal{M}
=\sum_{i,j}\wp^\star {}_i^{(j)}
=\sum_{j=1}^{m}\mathcal{Z}_jz^j.
\end{equation}
Consequently, the Helmholtz free energy per particle does not have only the usual contribution \(-\alpha/\beta\), but contains an additional term associated with the number of molecules,
\begin{equation}
\mathcal{F}
=U-\frac{1}{\beta}\mathcal{S}_{\mathrm{SF}}
=-\frac{\alpha}{\beta}-\frac{\mathcal{M}}{\beta}.
\end{equation}
This additional contribution is a direct thermodynamic signature of structure formation.

An instructive special case illustrates how structure-forming multiplicity alone can produce a nonuniform MaxEnt distribution. Suppose that each molecule size \(j=1,\ldots,m\) has only one possible state and that all energies vanish \(\epsilon^{(j)} \equiv 0\). Defining \(x=n\exp(-\alpha)\), the MaxEnt distribution becomes
\begin{equation}
\wp_j^\star=\frac{x^j}{n j!},
\end{equation}
while particle conservation determines \(x\) through the polynomial equation
\begin{equation}
x\sum_{k=0}^{m-1}\frac{x^k}{k!}=n.
\end{equation}
The probability that a randomly selected elementary particle belongs to a molecule of size \(j\) is therefore
\begin{equation}
\pi_j=j\wp_j^\star
=\frac{x^j}{n(j-1)!}.
\end{equation}

For fixed \(m\) and large \(n\), the highest-order term dominates, giving
\begin{equation}
x\sim\left[(m-1)!n\right]^{1/m},
\qquad
\pi_j\sim
\frac{[(m-1)!]^{j/m}}{(j-1)!}
n^{j/m-1}.
\end{equation}
Consequently, \(\pi_j\to0\) for every \(j<m\), whereas \(\pi_m\to1\). For example, \(m=2\) gives \(x\sim\sqrt{n}\), while \(m=3\) gives \(x\sim(2n)^{1/3}\); in both cases, particles become concentrated in the largest permitted molecules.

If arbitrarily large molecules are allowed, the finite sum can instead be approximated by an exponential. The normalization equation becomes
\begin{equation}
xe^x=n,
\qquad
x=\mathcal{W}(n),
\end{equation}
where \(\mathcal{W}\) is the Lambert function, and the particle-size distribution becomes a shifted Poisson distribution,
\begin{equation}
\pi_j=e^{-x}\frac{x^{j-1}}{(j-1)!}.
\end{equation}
Its typical structure size grows approximately as \(\mathcal{W}(n)\sim\ln n-\ln\ln n\). Thus, even when all structures have identical energies, the MaxEnt distribution is not uniform: it is determined by the combinatorial weights associated with structure formation.

Emergent structures can therefore change both entropy and equilibrium statistics even when the elementary particles remain distinguishable. The modified entropy follows from counting the allowed configurations rather than from postulating a generalized entropy. In the thermodynamic limit, the result is equivalent to a grand-canonical ensemble, but the two can differ substantially for finite systems \cite{Korbel2021natcom}. The same counting has also been used to study changes in the order of phase transitions caused by the presence of structure formation in systems \cite{Korbel2025natcom}.

\subsection{Sample-space reducing processes}

Sample-space reducing (SSR) processes describe path-dependent dynamics in which the set of states accessible at the next step depends on the current state and progressively contracts during relaxation, explicitly violating detailed balance. They provide simple models of dissipative cascades, fragmentation, directed diffusion, language production, and inelastic collisions, and naturally generate power-law distributions \cite{corominas2015pnas,corominas2017scirep,thurner2023njp}. Unlike the preceding examples, their relevant microstates are entire trajectories rather than static configurations. Consequently, their multiplicity counts the number of trajectories compatible with a given histogram of state visits.

Let us focus first on the simple example of slowly-driven SSR, where the system fully relaxes before a driving event. Consider an ordered state space
\begin{equation}
\mathbb{S}=\{s_1,s_2,\ldots,s_N\},
\end{equation}
where \(s_1\) is the lowest state. During an SSR relaxation step, a process in state \(s_i\) can move only to a lower state \(s_j\), with \(j<i\). Once the process reaches \(s_1\), it is restarted, and a new relaxation sequence begins. A trajectory microstate is represented by the ordered sequence
\begin{equation}
\boldsymbol{x}=(x_1,\ldots,x_n),
\qquad
x_t\in\mathbb{S},
\end{equation}
and can be decomposed into \(R\) decreasing sequences, or runs. Let \(n_i\) denote the total number of visits to state \(s_i\). The trajectory length is
\begin{equation}
n=\sum_{i=1}^{N}n_i.
\end{equation}
Since every run reaches \(s_1\) exactly once, \(n_1=R\). The visit histogram
\begin{equation}
\boldsymbol{n}=(n_1,\ldots,n_N)
\end{equation}
defines its mesostate.

To determine the multiplicity, consider a state \(s_i\) with \(i>1\). Because a decreasing run can visit \(s_i\) at most once, specifying which \(n_i\) of the \(R=n_1\) runs contain \(s_i\) completely determines its occurrence in the trajectory. These choices can be made independently for every state, because the order of all states within a run is fixed by the SSR condition. The number of trajectories with histogram \(\boldsymbol{n}\) is therefore
\begin{equation}
W_{\mathrm{SSR}}(\boldsymbol{n})
=
\prod_{i=2}^{N}\binom{n_1}{n_i}.
\end{equation}
This product of binomial coefficients replaces the multinomial multiplicity of independent sampling and explicitly incorporates the path constraint \cite{corominas2015pnas}. Introducing the visit frequencies
\begin{equation}
p_i=\frac{n_i}{n},
\qquad
\sum_{i=1}^{N}p_i=1,
\end{equation}
The quantity \(p_1=R/n\) is the number of relaxation runs per trajectory step. Applying Boltzmann's formula and Stirling's approximation gives the entropy per step
\begin{eqnarray}
\mathcal{S}_{\mathrm{SSR}}(\boldsymbol{p})
&=&
-\sum_{i=2}^{N}
\left[
p_i\ln\left(\frac{p_i}{p_1}\right)
\right.
\nonumber\\
&&\left.
+(p_1-p_i)
\ln\left(1-\frac{p_i}{p_1}\right)
\right].
\end{eqnarray}
The entropy is not Shannon entropy of the visit frequencies. Instead, every term is a binary entropy measuring whether a particular run visits state \(s_i\). This form follows directly from counting admissible decreasing trajectories and is an example of how path dependence changes the entropy entering the maximum-configuration principle \cite{Hanel2018entropy,hanel2023plos}.

The multiplicity specifies how many trajectories share the same histogram, but their probability also depends on the dynamical constraints. Let \(q_i\) be a normalized prior weight assigned to state \(s_i\), and define the cumulative weights
\begin{equation}
Q_i=\sum_{j=1}^{i}q_j.
\end{equation}
At a relaxation step from \(s_i\), the next state \(s_j\), \(j<i\), is sampled with the renormalized probability \(q_j/Q_{i-1}\). After reaching \(s_1\), the process is restarted by sampling from the unrestricted prior \(\boldsymbol{q}\). All trajectories with histogram \(\boldsymbol{n}\) then have the same probability
\begin{equation}
G_{\mathrm{SSR}}(\boldsymbol{n}\mid\boldsymbol{q})
=
\prod_{i=1}^{N}q_i^{n_i}
\prod_{i=2}^{N}Q_{i-1}^{-n_i}.
\end{equation}
The corresponding cross-entropy per step (equal to $\ln G$, similarly to Boltzmann's formula) is
\begin{equation}
\mathcal{C}_{\mathrm{SSR}}(\boldsymbol{p}\mid\boldsymbol{q})
=
-\sum_{i=1}^{N}p_i\ln q_i
+\sum_{i=2}^{N}p_i\ln Q_{i-1}.
\end{equation}
Thus, the prior weights and the restriction to nested sample spaces enter through the trajectory probability rather than through the multiplicity. For example, energetic information can be introduced through Boltzmann prior weights
\begin{equation}
q_i=\frac{\exp(-\beta\epsilon_i)}{\mathcal{Z}_N},
\qquad
\mathcal{Z}_i=\sum_{j=1}^{i}\exp(-\beta\epsilon_j),
\end{equation}
for which \(Q_i=\mathcal{Z}_i/\mathcal{Z}_N\). The cross-entropy then becomes
\begin{equation}
\mathcal{C}_{\mathrm{SSR}}(\boldsymbol{p}\mid\boldsymbol{q})
=
\beta U
+p_1\ln\mathcal{Z}_N
+\sum_{i=2}^{N}p_i\ln\mathcal{Z}_{i-1},
\end{equation}
where \(U=\sum_i p_i\epsilon_i\). It therefore contains the regular mean-energy contribution together with terms involving the partial partition functions of the nested accessible sample spaces.

The probability of observing a histogram is
\begin{equation}
P_{\mathrm{SSR}}(\boldsymbol{n}\mid\boldsymbol{q})
=
W_{\mathrm{SSR}}(\boldsymbol{n})
G_{\mathrm{SSR}}(\boldsymbol{n}\mid\boldsymbol{q}).
\end{equation}
In the long-trajectory limit, its most probable distribution maximizes
\begin{equation}
\mathcal{S}_{\mathrm{SSR}}(\boldsymbol{p})
-\mathcal{C}_{\mathrm{SSR}}(\boldsymbol{p}\mid\boldsymbol{q}).
\end{equation}
The stationarity conditions yield
\begin{equation}
p_i^\star
=p_1^\star\frac{q_i}{Q_i},
\qquad
p_1^\star
=
\left(\sum_{i=1}^{N}\frac{q_i}{Q_i}\right)^{-1}.
\end{equation}
This is the maximum-configuration distribution of the fully relaxing SSR process and provides the corresponding MaxEnt description. For uniform prior weights, \(q_i=1/N\) and \(Q_i=i/N\), it reduces to
\begin{equation}
p_i^\star
=\frac{1}{Z_N\,i},
\qquad
Z_N=\sum_{j=1}^{N}\frac{1}{j},
\end{equation}
which is Zipf's law. The power law, therefore, results from the shrinking accessible sample space rather than from imposing a power-law distribution as an external constraint \cite{corominas2015pnas}.

More general driven SSR processes need not complete a full relaxation run before being restarted. State-dependent driving rates allow the process to alternate between dissipative SSR steps and unrestricted driving steps, producing non-equilibrium steady states with explicitly broken detailed balance. Their multiplicities still count trajectories compatible with the observed histograms, but additional variables are required to record where driving events occur. The resulting entropy and cross-entropy separate into equilibrium-like Shannon contributions and correction terms that encode the broken detailed balance \cite{Hanel2018entropy,Hofer2025driven}. In the fast-driving limit, SSR relaxation becomes negligible, and the usual multinomial and Shannon expressions are recovered. Dynamical constraints modify the multiplicity, entropy, and corresponding MaxEnt distribution.

\subsection{P\'olya urns}

P\'olya urns are among the simplest models of reinforcement: observing a state increases the probability of observing the same state again. Like SSR processes, they are history-dependent, since the probability of the next event depends on the trajectory followed so far. However, the mechanism is different. Whereas SSR dynamics restricts the accessible sample space, P\'olya dynamics modifies the statistical weights of the states through reinforcement. The classical model was introduced by Eggenberger and P\'olya \cite{eggenberger1923zamm,polya1930quelques} and has become a basic example of non-ergodic and self-reinforcing dynamics \cite{pemantle2007probsurv}.

Consider an urn containing balls of \(N\) colors, represented by the state space
\begin{equation}
\mathbb{S}=\{s_1,\ldots,s_N\}.
\end{equation}
Initially, the urn contains \(a_i>0\) balls of color \(s_i\), with
\begin{equation}
A=\sum_{i=1}^{N}a_i.
\end{equation}
At every step, one ball is drawn and returned to the urn together with \(\delta>0\) additional balls of the same color. A trajectory microstate is the ordered sequence of observed colors,
\begin{equation}
\boldsymbol{x}=(x_1,\ldots,x_n),
\qquad
x_t\in\mathbb{S},
\end{equation}
and its mesostate is the histogram
\begin{equation}
\boldsymbol{n}=(n_1,\ldots,n_N),
\qquad
\sum_{i=1}^{N}n_i=n.
\end{equation}
After \(t\) draws, let \(n_i(t)\) denote the number of times that color \(s_i\) has been observed. The probability of observing \(s_i\) at the next step is
\begin{equation}
P(x_{t+1}=s_i\mid\boldsymbol{n}(t))
=
\frac{a_i+\delta n_i(t)}{A+\delta t}.
\end{equation}
Introducing the initial probabilities and the dimensionless reinforcement parameter
\begin{equation}
q_i=\frac{a_i}{A},
\qquad
\gamma=\frac{\delta}{A},
\end{equation}
This transition probability becomes
\begin{equation}
P(x_{t+1}=s_i\mid\boldsymbol{n}(t))
=
\frac{q_i+\gamma n_i(t)}{1+\gamma t}.
\end{equation}
In the limit \(\gamma\to0\), the process reduces to independent sampling from the fixed distribution \(\boldsymbol{q}\). For \(\gamma>0\), early fluctuations modify subsequent probabilities and can persist throughout the trajectory.

The number of ordered trajectories with a given histogram remains the multinomial coefficient
\begin{equation}
W(\boldsymbol{n})
=
\frac{n!}{\prod_{i=1}^{N}n_i!}.
\end{equation}
This does not mean that the process is multinomial: reinforcement changes the probability assigned to each trajectory. In the classical linear P\'olya urn, all trajectories sharing the same histogram have equal probability because the factors generated by successive observations can be reordered. Their common probability is
\begin{eqnarray}
G_{\mathrm{P}}(\boldsymbol{n}\mid\boldsymbol{q},\gamma)
&=&
\frac{\Gamma(1/\gamma)}
{\Gamma(n+1/\gamma)}
\nonumber\\
&&\times
\prod_{i=1}^{N}
\frac{\Gamma(n_i+q_i/\gamma)}
{\Gamma(q_i/\gamma)}.
\end{eqnarray}
Consequently, the probability of observing the mesostate \(\boldsymbol{n}\) is
\begin{equation}
P_{\mathrm{P}}(\boldsymbol{n}\mid\boldsymbol{q},\gamma)
=
W(\boldsymbol{n})
G_{\mathrm{P}}(\boldsymbol{n}\mid\boldsymbol{q},\gamma),
\end{equation}
which is the Dirichlet--multinomial distribution \cite{thurner2017three}.

This factorization illustrates an important difference from the previous examples. The multinomial multiplicity alone produces the Shannon entropy, but its leading contribution is canceled by an opposing contribution from the reinforcement-dependent trajectory probability. Writing \(p_i=n_i/n\), the part of the logarithmic histogram probability that depends on \(\boldsymbol{p}\) is, for large \(n\),
\begin{equation}
\ln P(\boldsymbol{p}\mid\boldsymbol{q},\gamma)
\simeq
\sum_{i=1}^{N}
\left(\frac{q_i}{\gamma}-1\right)\ln p_i
+\mathrm{const}.
\end{equation}
Following the maximum-configuration factorization, this expression can be separated into the entropy and cross-entropy terms \cite{thurner2017three}
\begin{equation}
\mathcal{S}_{\mathrm{P}}(\boldsymbol{p})
=-\sum_{i=1}^{N}\ln p_i,
\qquad
\mathcal{C}_{\mathrm{P}}(\boldsymbol{p}\mid\boldsymbol{q})
=-\frac{1}{\gamma}\sum_{i=1}^{N}q_i\ln p_i,
\end{equation}
such that the most probable distribution maximizes
\begin{equation}
\mathcal{S}_{\mathrm{P}}(\boldsymbol{p})
-\mathcal{C}_{\mathrm{P}}(\boldsymbol{p}\mid\boldsymbol{q}).
\end{equation}
The entropy \(\mathcal{S}_{\mathrm{P}}\) is the negative of the Burg entropy introduced earlier. At finite \(n\), the logarithms are regularized as \(\ln(p_i+1/n)\), so configurations on the boundary remain well defined \cite{thurner2017three}.

If \(q_i>\gamma\) for every state, the maximum lies in the interior of the probability simplex. Maximizing the asymptotic histogram probability under normalization gives
\begin{equation}
p_i^\star
=
\frac{q_i-\gamma}{1-N\gamma}.
\end{equation}
For weak reinforcement, \(\gamma\to0\), this approaches the initial distribution \(p_i^\star=q_i\), consistently recovering Bernoulli sampling. Nevertheless, the classical linear P\'olya process is non-ergodic: the empirical frequencies of an individual trajectory do not converge to one deterministic distribution. Instead, their limiting distribution is Dirichlet, and early random events determine where an individual realization ends \cite{pemantle2007probsurv}. The interior maximum therefore describes the most probable frequency distribution across an ensemble of urn realizations, but fluctuations around it do not vanish in the usual thermodynamic-limit sense.

When reinforcement is sufficiently strong, one or more coefficients \(q_i/\gamma-1\) become negative and the maximum moves toward the boundary of the simplex. In nonlinear P\'olya urns, where the probability of drawing a color grows faster than linearly with its abundance, this instability can lead to a genuine winner-takes-all regime: an early advantage is amplified until one state is sampled almost exclusively \cite{pemantle2007probsurv,thurner2017three}. The negative Burg form makes this boundary preference transparent, since maximizing \(-\sum_i\ln p_i\) favors distributions containing very small probabilities. This behavior contrasts with the ordinary Burg entropy, \(\sum_i\ln p_i\), which is concave and favors an interior distribution.

P\'olya-like models can also describe systems in which reinforcement is accompanied by the appearance of new possibilities. In urn models with triggering, drawing a previously unseen color introduces additional colors into the urn, so the accessible state space expands whenever a novelty occurs. Such models formalize the exploration of the \emph{adjacent possible} and connect reinforcement and path dependence with the emergence of Heaps', Zipf's, and Taylor's laws \cite{tria2014scirep,tria2018entropy,loreto2026chapter}. In these models, history dependence changes both the weights of existing states and the set of states that can be visited in the future.

\subsection{Tsallis entropy and q-exponentials}

The preceding examples produced non-Shannon entropies by changing either the microscopic counting rules or the probability assigned to trajectories. It is also possible to construct a generalized entropy by deforming the mathematical operations used in Boltzmann's formula itself. A prominent example is the combinatorial construction of Tsallis entropy through \(q\)-deformed algebra \cite{nivanen2003rmp,borges2004physa,suyari2006physa}. As we will see, this establishes a formal analogy with the derivation of Shannon entropy from the multinomial coefficient, but requires replacing the ordinary logarithm, product, factorial, multinomial coefficient, and Stirling approximation by their deformed counterparts.

The \(q\)-logarithm and \(q\)-exponential are defined as
\begin{equation}
\ln_q(x)
=
\frac{x^{1-q}-1}{1-q},
\qquad
\exp_q(x)
=
\left[1+(1-q)x\right]_+^{1/(1-q)},
\end{equation}
where \([x]_+=\max(x,0)\). They are mutually inverse and recover the ordinary logarithm and exponential in the limit \(q\to1\). The \(q\)-logarithm satisfies the pseudo-additive composition rule
\begin{equation}
\ln_q(xy)
=
\ln_qx+\ln_qy
+(1-q)\ln_qx\ln_qy,
\end{equation}
which mirrors the pseudo-additivity of Tsallis entropy.

To restore an additivity of logarithm, one introduces the \(q\)-product \cite{nivanen2003rmp,borges2004physa}
\begin{equation}
x\otimes_q y
=
\left[x^{1-q}+y^{1-q}-1\right]_+^{1/(1-q)}.
\end{equation}
By construction,
\begin{equation}
\ln_q(x\otimes_q y)
=
\ln_qx+\ln_qy.
\end{equation}
The corresponding \(q\)-division is
\begin{equation}
x\oslash_q y
=
\left[x^{1-q}-y^{1-q}+1\right]_+^{1/(1-q)},
\end{equation}
and satisfies
\begin{equation}
\ln_q(x\oslash_q y)
=
\ln_qx-\ln_qy.
\end{equation}
Both operations reduce to ordinary multiplication and division as \(q\to1\). They are designed to linearize the \(q\)-logarithm, but do not preserve all properties of ordinary arithmetic; in particular, the \(q\)-product is generally not distributive over ordinary addition.
The following construction is understood on the domain where the deformed operations and their intermediate results remain well defined; depending on \(q\), the positive-part cutoff can otherwise restrict the admissible arguments.

Using the \(q\)-product, the \(q\)-factorial is defined recursively as
\begin{equation}
n!_q
=
1\otimes_q2\otimes_q\cdots\otimes_q n,
\end{equation}
and the \(q\)-multinomial coefficient becomes
\begin{equation}
\binom{n}{n_1,\ldots,n_N}_q
=
n!_q
\oslash_q
\left(n_1!_q\otimes_q\cdots\otimes_q n_N!_q\right),
\end{equation}
where \(\sum_i n_i=n\). Applying the \(q\)-logarithm turns this deformed quotient into a difference,
\begin{equation}
\ln_q
\binom{n}{n_1,\ldots,n_N}_q
=
\ln_q(n!_q)
-\sum_{i=1}^{N}\ln_q(n_i!_q).
\end{equation}
This has the same algebraic structure as the logarithm of the ordinary multinomial coefficient, but its constituent operations have all been deformed.

The generalized counterpart of Stirling's approximation follows from
\begin{equation}
\ln_q(n!_q)
=
\sum_{k=1}^{n}\ln_q k.
\end{equation}
Approximating the sum by an integral gives, for \(q\neq2\),
\begin{equation}
\ln_q(n!_q)
\simeq
\frac{n}{2-q}\ln_q n
-\frac{n}{2-q}.
\end{equation}
Using this \(q\)-Stirling approximation in the \(q\)-multinomial coefficient and introducing \(p_i=n_i/n\) yields \cite{suyari2006physa}
\begin{equation}
\ln_q
\binom{n}{n_1,\ldots,n_N}_q
\simeq
\frac{n^{2-q}}{2-q}
S_{2-q}(\boldsymbol{p}),
\end{equation}
where
\begin{equation}
S_{2-q}(\boldsymbol{p})
=
\frac{\sum_i p_i^{2-q}-1}{q-1}.
\end{equation}
Equivalently, after replacing \(q\) by its additive dual \(2-q\), one obtains
\begin{equation}
\ln_{2-q}
\binom{n}{n_1,\ldots,n_N}_{2-q}
\simeq
\frac{n^q}{q}
S_q(\boldsymbol{p}).
\end{equation}
This relation is the deformed analogue of
\begin{equation}
\ln\binom{n}{n_1,\ldots,n_N}
\simeq
nH(\boldsymbol{p}),
\end{equation}
and reduces to it in the limit \(q\to1\).

The construction therefore relates Tsallis entropy to a generalized Boltzmann formula by replacing the ordinary multinomial coefficient and logarithm with their \(q\)-deformed counterparts. Its interpretation is complementary to the preceding examples: rather than beginning with specific microscopic configurations or trajectory probabilities, it begins with an algebra adapted to the composition properties of Tsallis entropy.

This approach reveals a possible combinatorial structure for Tsallis statistics, although it introduces more mathematical machinery than ordinary microscopic counting. Its physical interpretation depends on whether the \(q\)-deformed operations can be connected to the dynamics of the system under consideration.

\section{Axiomatic approaches}

In the previous section, entropy was derived from the microscopic structure and dynamics of particular systems. Here, we instead begin with properties that an entropy should satisfy and ask which entropy functionals meet them. The frameworks below address coarse-graining, the composition of independent systems, the scaling of phase-space volume, the consistency of statistical inference, and the phenomenological structure of thermodynamics.

\subsection{Shannon-Khinchin axioms}

The Shannon--Khinchin (SK) axioms characterize entropy as a measure of uncertainty associated with a discrete probability distribution. Let \(A\) be a random variable with possible outcomes \(\{A_i\}_{i=1}^{N}\) and probability distribution \(\boldsymbol{p}=(p_1,\ldots,p_N)\). Its entropy is denoted by \(S(A)=S(\boldsymbol{p})\). The original SK axioms can be stated as follows \cite{khinchin2013mathematical}:

\paragraph{SK1: Continuity.}
The entropy is a continuous function of all probabilities,
\begin{equation}
S:\Delta_N\longrightarrow\mathbb{R}
\quad\text{is continuous},
\end{equation}
where \(\Delta_N\) denotes the probability simplex.
Small changes in the probability distribution therefore produce small changes in the associated uncertainty.

\paragraph{SK2: Maximality.}
For a fixed number \(N\) of possible outcomes, entropy is maximal for the uniform distribution \(\boldsymbol{u}_N=(1/N,\ldots,1/N)\),
\begin{equation}
\max_{\boldsymbol{p}}S(\boldsymbol{p})=S(\boldsymbol{u}_N).
\end{equation}
Thus, uncertainty is greatest when all outcomes are equally probable.

\paragraph{SK3: Expandability.}
Adding an impossible outcome does not change the entropy,
\begin{equation}
S(p_1,\ldots,p_N,0)=S(p_1,\ldots,p_N).
\end{equation}

\paragraph{SK4\(_{\mathrm S}\): Shannon additivity.}
For two random variables \(A\) and \(B\), the joint entropy satisfies the chain rule
\begin{equation}
S(A,B)
=
S(B\mid A)+S(A)
=
S(A\mid B)+S(B),
\end{equation}
where the conditional entropy is the ordinary probability-weighted average
\begin{equation}
S(B\mid A)
=
\sum_{i=1}^{N}p_i S(B\mid A=A_i).
\end{equation}
This axiom ensures that uncertainty can be evaluated consistently either jointly or successively. In particular, if \(A\) and \(B\) are statistically independent, then \(S(B\mid A)=S(B)\), and the chain rule reduces to ordinary additivity,
\begin{equation}
S(A,B)=S(A)+S(B).
\end{equation}
Together, SK1--SK4\(_{\mathrm S}\) uniquely determine the entropy up to a positive multiplicative constant,
\begin{equation}
S(\boldsymbol{p})
=
K H(\boldsymbol{p})
=
-K\sum_{i=1}^{N}p_i\ln p_i,
\qquad
K>0.
\end{equation}

Many generalized entropies preserve SK1--SK3 and modify only the fourth axiom, which determines how entropy behaves under conditional decomposition and composition. Two important examples are provided by R\'enyi and Tsallis entropy. In the axiomatizations presented below, their parameter is restricted to \(q>0\).

\paragraph{SK4\(_{\mathrm R}\): R\'enyi additivity.}
R\'enyi entropy preserves the ordinary additive chain rule \cite{renyi1961proceedings},
\begin{equation}
R_q(A,B)
=
R_q(B\mid A)+R_q(A),
\end{equation}
but replaces the linear average defining conditional entropy by an escort-weighted Kolmogorov--Nagumo mean,
\begin{equation}
R_q(B\mid A)
=
f^{-1}\left(
\sum_i\rho_i(q)
f\left(R_q(B\mid A=A_i)\right)
\right),
\end{equation}
where
\begin{equation}
\rho_i(q)
=
\frac{p_i^q}{\sum_jp_j^q}
\end{equation}
is the escort distribution. The axioms restrict the KN function to an exponential form for \(q\neq1\), yielding
\begin{equation}
R_q(\boldsymbol{p})
=
\frac{1}{1-q}
\ln\left(\sum_i p_i^q\right).
\end{equation}
Thus, R\'enyi entropy remains additive for independent systems while modifying how conditional entropies are averaged. Kolmogorov--Nagumo means, and their thermodynamic implications will be discussed in more detail later in this chapter.

\paragraph{SK4\(_{\mathrm T}\): Tsallis additivity.}
For Tsallis entropy, conditional entropy is defined through an escort-weighted linear average,
\begin{equation}
S_q(B\mid A)
=
\sum_i\rho_i(q)S_q(B\mid A=A_i),
\end{equation}
while the chain rule is replaced by the pseudo-additive composition law \cite{abe2000physletta,suyari2004ieee}
\begin{eqnarray}
S_q(A,B)
&=&
S_q(B\mid A)+S_q(A)
\nonumber\\
&&
+(1-q)S_q(B\mid A)S_q(A).
\end{eqnarray}
Together with SK1--SK3, this modified axiom uniquely determines the Tsallis--Havrda--Charv\'at entropy,
\begin{equation}
S_q(\boldsymbol{p})
=
\frac{\sum_i p_i^q-1}{1-q}.
\end{equation}
For statistically independent variables, the conditional entropy reduces to \(S_q(B)\), and the generalized chain rule becomes the familiar pseudo-additive composition rule. Both SK4\(_{\mathrm R}\) and SK4\(_{\mathrm T}\) reduce to SK4\(_{\mathrm S}\) as \(q\to1\).

More generally, SK4 can be formulated as a composability requirement,
\begin{equation}
S(A,B)
=
f(S(B\mid A), S(A))
\end{equation}
where \(f\) denotes a generalized composition operation and the conditional entropy may itself involve a generalized average. This includes broad classes of strongly pseudo-additive entropies, such as Sharma--Mittal-type families, obtained by combining generalized conditional averaging with non-additive composition rules \cite{Ilic2014physa,Sharma1975metrika,Mittal1975metrika}. Different thermostatistical formulations based on strongly pseudo-additive entropies can also be related through suitable transformations of their entropy, constraints, and Lagrange multipliers \cite{Ilic2010pre}. The fourth axiom therefore plays the key role: SK1--SK3 specify basic regularity properties, whereas SK4 determines how information is assembled from subsystems and successive observations. The group-composability framework discussed next treats admissible composition rules in more detail.

\subsection{Tempesta group composability}

The generalized fourth Shannon--Khinchin axiom introduced above allows the entropy of a composite system to be determined through a generalized composition function. Tempesta's group-composability framework systematically investigates which composition laws are mathematically consistent and which entropy functionals realize them \cite{Tempesta2011pre,tempesta2016annals}. For two statistically independent systems \(A\) and \(B\), composability requires
\begin{equation}
S(A,B)
=
\Phi\left(S(A),S(B)\right),
\end{equation}
where the entropy of the combined system depends only on the entropies of its components.

A consistent composition law should be symmetric,
\begin{equation}
\Phi(x,y)=\Phi(y,x),
\end{equation}
associative,
\begin{equation}
\Phi\left(\Phi(x,y),z\right)
=
\Phi\left(x,\Phi(y,z)\right),
\end{equation}
and possess a neutral element,
\begin{equation}
\Phi(x,0)=x.
\end{equation}
Symmetry reflects the equivalence of exchanging independent subsystems, associativity ensures that the entropy of several systems does not depend on how they are grouped, and the neutral element represents composition with a system of vanishing entropy. Under suitable regularity conditions, these requirements identify \(\Phi\) with a commutative formal group law.

The ordinary additive law,
\begin{equation}
\Phi_{\mathrm{add}}(x,y)=x+y,
\end{equation}
describes both Shannon and R\'enyi entropy for independent systems. Tsallis entropy instead follows the deformed law
\begin{equation}
\Phi_q(x,y)
=
x+y+(1-q)xy.
\end{equation}
Despite its non-additive form, this rule is symmetric, associative, and has zero as its neutral element. The pseudo-additivity of Tsallis entropy can therefore be understood as a particular formal group law rather than as a failure of consistent composition.

Locally, a commutative one-dimensional formal group law can be expressed through a group exponential \(G\) as
\begin{equation}
\Phi(x,y)
=
G\left(G^{-1}(x)+G^{-1}(y)\right).
\end{equation}
The inverse function \(G^{-1}\) is the corresponding group logarithm. It maps the generalized composition law to ordinary addition,
\begin{equation}
G^{-1}\left(\Phi(x,y)\right)
=
G^{-1}(x)+G^{-1}(y).
\end{equation}
For the additive law, \(G(t)=t\). For Tsallis composition, one may choose
\begin{equation}
G_q(t)
=
\frac{\exp[(1-q)t]-1}{1-q},
\end{equation}
which satisfies \(G_q(x+y)=\Phi_q(G_q(x),G_q(y))\).

The formal-group construction motivates the trace-form group entropy
\begin{equation}
S_G(\boldsymbol{p})
=
\sum_{i=1}^{N}
p_i G\left(\ln\frac{1}{p_i}\right).
\end{equation}
The choice \(G(t)=t\) gives Shannon entropy, whereas \(G=G_q\) gives
\begin{equation}
S_G(\boldsymbol{p})
=
\frac{\sum_i p_i^q-1}{1-q},
\end{equation}
which is Tsallis entropy. More general choices of the group exponential generate broad families of generalized entropies. The universal-group entropy provides a parametrization of such trace-form measures in terms of the coefficients of a general formal group exponential \cite{tempesta2016annals}.

An important distinction is made between \emph{weak} and \emph{strong} composability. An entropy is weakly composable when its composition law holds for uniform probability distributions. This ensures a consistent rule for combining maximally uncertain systems but does not necessarily determine the entropy of arbitrary composite distributions. Strong composability requires
\begin{equation}
S(\boldsymbol{p}^{A}\boldsymbol{p}^{B})
=
\Phi\left(S(\boldsymbol{p}^{A}),S(\boldsymbol{p}^{B})\right)
\end{equation}
for arbitrary independent distributions, where
\begin{equation}
\left(\boldsymbol{p}^{A}\boldsymbol{p}^{B}\right)_{ij}
=p_i^{A}p_j^{B}.
\end{equation}
Strong composability is considerably more restrictive. Shannon and Tsallis entropy satisfy it, while a general trace-form group entropy is typically only weakly composable.

The formal-group approach can also be extended beyond trace-form entropies. The resulting \(Z\)-entropies form a broad class of strongly composable measures and include R\'enyi entropy as a particular example \cite{tempesta2017rspa}. Group composability classifies entropy functionals by how information combines across independent systems. The Hanel--Thurner approach discussed next instead classifies them by their asymptotic scaling with the size of the accessible phase space.

\subsection{Hanel-Thurner asymptotic scaling}

The entropy appropriate for a statistical system is closely related to the growth of its phase-space volume. Let \(W_{\mathrm{tot}}(n)\) denote the total number of accessible configurations of a system of size \(n\), distinguishing it from the multiplicity \(W(\boldsymbol{n})\) of a particular mesostate used earlier. For the uniform distribution \(\boldsymbol{u}_{W_{\mathrm{tot}}(n)}\), an entropy is \emph{extensive} if
\begin{equation}
S\left(\boldsymbol{u}_{W_{\mathrm{tot}}(n)}\right)
\sim n
\end{equation}
in the thermodynamic limit. Shannon entropy satisfies
\begin{equation}
H\left(\boldsymbol{u}_{W_{\mathrm{tot}}(n)}\right)=\ln W_{\mathrm{tot}}(n)
\end{equation}
and is therefore extensive when \(W_{\mathrm{tot}}(n)\) grows exponentially. Many complex systems may instead possess sub-exponentially or super-exponentially growing phase spaces, in which case another entropy may be required to restore extensivity. The Hanel--Thurner framework classifies systems and their extensive entropies through the asymptotic scaling of \(W_{\mathrm{tot}}(n)\) \cite{Hanel2011epl1,Hanel2011epl2}.

The original classification focuses on trace-form entropies
\begin{equation}
S_g(\boldsymbol{p})
=
\sum_{i=1}^{N}g(p_i),
\end{equation}
where \(g\) is continuous, concave, and satisfies \(g(0)=0\). Their leading asymptotic behavior for small probabilities is characterized by two scaling exponents \(c\) and \(d\). The first exponent is defined through
\begin{equation}
\lim_{x\to0^+}
\frac{g(zx)}{g(x)}
=z^c,
\end{equation}
whereas the second exponent follows from
\begin{equation}
\lim_{x\to0^+}
\frac{g(x^{1+a})}{x^{ac}g(x)}
=(1+a)^d.
\end{equation}
Entropies sharing the same exponents \((c,d)\) belong to the same asymptotic equivalence class. Their detailed finite-size forms can differ, but their leading thermodynamic behavior is identical.

The first three Shannon--Khinchin axioms restrict the admissible parameter space. Continuity and expandability require \(c>0\), while maximality, which follows from concavity of \(g\), implies \(c\leq1\). Thus,
\begin{equation}
0<c\leq1,
\end{equation}
whereas \(d\) can take any real value. The limiting case \(c=1\) additionally requires \(d\geq0\) for concavity. These restrictions determine which scaling classes can be represented by admissible trace-form entropies \cite{Hanel2011epl1,Hanel2011epl2}.

For Shannon entropy, \(g(x)=-x\ln x\), and the scaling exponents are
\begin{equation}
(c,d)_{\mathrm{Shannon}}=(1,1).
\end{equation}
Tsallis entropy with \(0<q<1\) belongs to the class
\begin{equation}
(c,d)_{\mathrm{Tsallis}}=(q,0),
\end{equation}
while stretched-exponential phase spaces are associated with classes of the form \((1,d)\). The pair \((c,d)\) therefore provides a compact classification of broad families of generalized entropies and the sample-space growth laws for which they become extensive.

This relation can be understood directly through the phase-space volume. Let \(\Lambda\) be a generalized logarithm chosen as the asymptotic inverse of \(W_{\mathrm{tot}}\),
\begin{equation}
\Lambda(W_{\mathrm{tot}}(n))\sim n.
\end{equation}
The associated trace-form entropy
\begin{equation}
S_{\Lambda}(\boldsymbol{p})
=
\sum_i p_i\Lambda\left(\frac{1}{p_i}\right)
\end{equation}
satisfies
\begin{equation}
S_{\Lambda}\left(\boldsymbol{u}_{W_{\mathrm{tot}}(n)}\right)
=
\Lambda(W_{\mathrm{tot}}(n))
\sim n.
\end{equation}
Thus, the extensive entropy can be constructed from the asymptotic inverse of the phase-space growth function.

The \((c,d)\)-classification captures many important sub-exponential and exponential growth laws, but the diversity of complex systems requires a more general hierarchy. The scaling-expansion framework extends the classification to arbitrary sample-space growth, including superexponential systems \cite{Korbel2018njp}. To avoid the domain problems of ordinary nested logarithms, we introduce regularized iterated logarithms
\begin{equation}
\ln^{(0)}x=x,
\qquad
\ln^{(k+1)}x=1+\ln\left(\ln^{(k)}x\right).
\end{equation}
They satisfy \(\ln^{(k)}x\geq1\) for \(x\geq1\) and have the same leading asymptotic behavior as ordinary nested logarithms. We define \(\exp^{(k)}\) as their inverse functions. The asymptotic growth of the sample space can then be represented through a scaling expansion of the form
\begin{equation}
W_{\mathrm{tot}}(n)
\sim
\exp^{(l)}
\left[
\prod_{j=0}^{k}
\left(\ln^{(j)}n\right)^{c_j}
\right].
\end{equation}
Here, the integer \(l\) specifies the exponential order of the sample-space growth, while the sequence of scaling exponents \(\{c_j\}\) determines its universality class. Only the first exponents that do not vanish are needed to distinguish the leading and successive subleading growth orders.

The corresponding extensive entropy is obtained by asymptotically inverting the scaling expansion of \(W_{\mathrm{tot}}(n)\). For a sample space with exponential order \(l\) and scaling exponents \(\{c_j\}\), a generalized logarithm that is well defined for every \(x\geq1\) can be written as \cite{Korbel2020epjst}
\begin{equation}
\Lambda_{\{d_j\}}(x)
=
r
\left\{
\left[\ln^{(l)}x\right]^{d_0}
\prod_{j=1}^{k}
\left[\ln^{(j+l)}x\right]^{d_j}
-1
\right\},
\end{equation}
where the entropy exponents are related to the sample-space exponents by
\begin{equation}
d_0=\frac{1}{c_0},
\qquad
d_j=-\frac{c_j}{c_0},
\quad j\geq1.
\end{equation}
Here, \(r>0\) is a normalization constant. The generalized logarithm satisfies
\begin{equation}
\Lambda_{\{d_j\}}(1)=0.
\end{equation}

Inserting the scaling expansion of \(W_{\mathrm{tot}}(n)\) gives
\begin{equation}
\Lambda_{\{d_j\}}(W_{\mathrm{tot}}(n))\sim n,
\end{equation}
up to asymptotically irrelevant multiplicative constants. An extensive trace-form entropy for this universality class is therefore
\begin{equation}
S_{\{d_j\}}(\boldsymbol{p})
=
\sum_i p_i
\Lambda_{\{d_j\}}\left(\frac{1}{p_i}\right).
\end{equation}
The regularized iterated logarithms ensure that the entropy is defined throughout the probability simplex; admissible choices of the exponents must additionally preserve concavity.

The following examples illustrate how different growth laws determine different extensive entropies.

\paragraph{Polynomial growth.}
The simplest sub-exponential class has
\begin{equation}
W_{\mathrm{tot}}(n)\sim n^\alpha.
\end{equation}
It can arise when the effective number of independent updates grows only logarithmically with the observation time. A simple aging random walk, whose rate of taking new steps decreases with its age, explores only a polynomial number of trajectories during a time interval \(n\). Such systems are strongly constrained relative to an ordinary random walk, whose number of trajectories grows exponentially \cite{Korbel2018njp}.

\paragraph{Stretched-exponential growth.}
More generally, aging processes can produce
\begin{equation}
W_{\mathrm{tot}}(n)\sim\exp(n^\gamma),
\qquad
0<\gamma<1.
\end{equation}
For an aging random walk, this occurs when the number of effective steps grows as \(n^\gamma\). Each effective step still contributes an exponential number of alternatives, but the sublinear operational time reduces the total sample-space growth to a stretched exponential \cite{Korbel2018njp}.

\paragraph{Exponential growth.}
Independent or weakly correlated constituents typically generate
\begin{equation}
W_{\mathrm{tot}}(n)\sim\exp(\kappa n).
\end{equation}
Examples include sequences of independent coin tosses and systems with multinomial statistics. In this class, \(l=1\), \(c_0=1\), and \(d_0=1\), so the extensive generalized logarithm is asymptotically equivalent to \(\ln x\), recovering Shannon entropy.

\paragraph{Super-exponential growth.}
When a system can form new structures from its constituents, the number of configurations may grow faster than exponentially. The magnetic-coins model combines ordinary spin configurations with the possible ways of grouping coins into interacting structures and produces growth of the form
\begin{equation}
W_{\mathrm{tot}}(n)\sim n^n=\exp(n\ln n)
\end{equation}
up to subleading factors \cite{Jensen2018jphysa,Korbel2018njp}. This is closely related to the structure-forming systems discussed earlier.

\paragraph{Double-exponential growth.}
The scaling hierarchy also includes sample spaces whose exponential order exceeds one. Consider a cascading random walk in which every walker can move left, move right, or split into two independent walkers at each step. Because splitting increases the number of walkers that branch independently at subsequent steps, the number of possible histories grows approximately as
\begin{equation}
W_{\mathrm{tot}}(n)\sim\exp(\exp(n)).
\end{equation}
Such a process belongs to a double-exponential class with \(l=2\), and its extensive generalized logarithm is asymptotically equivalent to \(\ln\ln x\) \cite{Korbel2018njp}.

Systems with the same exponential order and scaling exponents possess asymptotically equivalent generalized logarithms and therefore belong to the same universality class. This provides a systematic extension of the original \((c,d)\)-classification: the latter retains the first two relevant scaling exponents, while the complete expansion can distinguish more complicated sample spaces and include super-exponential growth.

The scaling approach classifies generalized entropies by the asymptotic growth of the accessible configuration space. Composability asks how entropy combines across independent systems, whereas scaling asks which entropy becomes extensive for a given growth of phase-space volume. The scaling exponents also define distinct information-geometric structures that further characterize the corresponding universality classes \cite{Korbel2020epjst}.

\subsection{Shore-Johnson axioms}

The Shannon--Khinchin axioms characterize entropy as a measure of information, whereas the Shore--Johnson (SJ) axioms characterize the consistency of the inference procedure based on entropy maximization. In the discrete setting, one seeks the probability distribution
\begin{equation}
\boldsymbol{p}^{\star}
=
\arg\max_{\boldsymbol{p}\in\mathcal{C}}
S(\boldsymbol{p}),
\end{equation}
where the constraint set \(\mathcal{C}\) represents the available information. In the Shore--Johnson framework, this information is expressed through normalization and linear expectation values,
\begin{equation}
\sum_{i=1}^{N}p_i=1,
\qquad
\sum_{i=1}^{N}p_i I_{i,k}
=
\overline{I}_k,
\quad k=1,\ldots,m.
\end{equation}
The axioms require that equivalent formulations and decompositions of this inference problem produce consistent estimates of inferred probability distributions \cite{Shore1980ieee,Shore1981ieee}. More generally, extending the analysis for nonuniform priors can be formulated by minimizing a cross-entropy or divergence, but the discrete MaxEnt formulation is sufficient for the discussion below. 

The Shore-Johnson axioms were originally formulated as follows:

\paragraph{SJ1: Uniqueness.}
\emph{The result should be unique.} 

\noindent The inference procedure should yield one unique solution. If several distributions satisfy the constraints equally well, the procedure would not provide an unambiguous update of the available information.

\paragraph{SJ2: Permutation invariance.}
\emph{The result should be invariant under a change of coordinates.} 

\noindent In the discrete setting, this means that relabeling the states must only relabel the inferred probabilities. For any permutation \(\sigma\),
\begin{equation}
S(p_{\sigma(1)},\ldots,p_{\sigma(N)})
=
S(p_1,\ldots,p_N).
\end{equation}
The inference must therefore depend on the probabilities and constraints rather than on arbitrary state labels.

\paragraph{SJ3: Subset independence.}
\emph{It should not matter whether one treats independent subsets separately or together.} 

\noindent If constraints refer only to particular subsets of states, inference within one subset should not depend on whether unrelated subsets are treated separately or simultaneously, or whether the problem is formulated using conditional distributions or the full joint distribution. No information about one subset should be introduced merely through the way the full problem is partitioned.

\paragraph{SJ4: System independence.}
\emph{It should not matter whether one treats independent systems separately or together.} 

\noindent Suppose two systems \(A\) and \(B\) are subject to independent constraints. Inferring their joint distribution and then taking its marginals should give the same marginal distributions as inferring each system separately. This requirement expresses consistency of inductive inference without necessarily prescribing the complete correlation structure of the joint distribution.

The precise interpretation of system independence is central to the scope of the Shore--Johnson axioms. The original formulation requires consistency of the inferred marginal distributions and permits a broad class of entropic functionals \cite{Shore1980ieee,uffink1995shpsb}. A stronger formulation additionally requires the inferred joint distribution to factorize,
\begin{equation}
p_{ij}^{A+B}=p_i^A p_j^B.
\end{equation}
This product-factorization condition is more restrictive and is closely associated with the properties studied in the subsequent cross-entropy formulation \cite{Shore1981ieee}. It is appropriate when independence of the full systems is part of the supplied information, but it should not be inferred solely from independent marginal constraints.

Constraints imposed on a composite system also require care because they need not decompose into independent constraints on its subsystems. Their formulation can therefore determine whether a joint MaxEnt problem is equivalent to separate subsystem optimizations and how the resulting subsystem distributions are related \cite{Ramshaw2022pre}.

Uffink showed that the original consistency requirements permit a broad family of entropy functionals rather than selecting Shannon entropy alone \cite{uffink1995shpsb}. In the discrete setting, this Uffink class can be represented as
\begin{equation}
S_q^f(\boldsymbol{p})
=
f\left[
\left(\sum_i p_i^q\right)^{1/(1-q)}
\right],
\end{equation}
where \(f\) is a strictly increasing function and \(q>0\), with \(q=1\) understood through the continuous limit. Because monotonic transformations preserve the ordering of probability distributions, all choices of \(f\) for fixed \(q\) yield the same MaxEnt distribution under the same linear constraints. Suitable choices recover R\'enyi, Tsallis, and related entropy representations.

The compatibility of non-Shannon entropies with the Shore--Johnson axioms became the subject of an extended debate. Press\'e and collaborators argued that nonadditive entropies can introduce biases not warranted by the supplied data and interpreted system independence through product factorization \cite{Presse2013prl}. Tsallis questioned whether this interpretation of the axioms is suitable for strongly correlated complex systems \cite{Tsallis2015entropy}, while Press\'e and collaborators defended their formulation \cite{Presse2015entropy}.

It was shown that the original Shore--Johnson consistency requirements allow the full Uffink class and do not require Shannon entropy alone \cite{Jizba2019prl}. The stronger factorization condition supplies additional information about the absence of correlations and therefore should not be imposed when only the marginal constraints are known. The study illustrates the role of weak correlations and discusses applications to two-qubit quantum systems and strongly interacting nuclear systems, where non-Shannonian entropies can arise consistently within statistical inference.

The relation between information-theoretic and inference-based axioms was subsequently made explicit by reformulating the Shannon--Khinchin framework through generalized arithmetic \cite{Jizba2020pre}. For a strictly monotonic function \(f\), define the generalized product
\begin{equation}
x\otimes_f y
=
f\left(f^{-1}(x) \cdot f^{-1}(y)\right).
\end{equation}
Because  \(D_q(\boldsymbol{p})
= \left(\sum_i p_i^q\right)^{1/(1-q)}
\) is multiplicative for independent systems, the Uffink entropy \(S_q^f(\boldsymbol{p})=f(D_q(\boldsymbol{p}))\) satisfies
\begin{equation}
S_q^f(A,B)
=
S_q^f(B\mid A)\otimes_f S_q^f(A).
\end{equation}
This generalized product is equivalent, after a monotonic change of entropy scale, to the generalized pseudo-additive composition laws introduced in the modified SK4 axiom. Consequently, the generalized SK and original SJ axioms generate the same Uffink class of entropies. The two approaches express the same consistency structure from different viewpoints: SK axioms describe how a system's entropy is constructed from its subsystems, while SJ axioms specify how probability distributions should be inferred from partial information.

A related question is how the parameter of a generalized entropy (e.g., Tsallis nonextensivity parameter \(q\)) should itself be selected from data. Somazzi and Garlaschelli proposed the additional requirement that no entropic parameter should be inferable from a completely uninformative uniform distribution \cite{Somazzi2023arxiv}. Their approach connects generalized MaxEnt inference with maximum likelihood and provides a data-driven criterion for identifying admissible entropy parameters.

Whereas the preceding frameworks characterize entropy through information and consistent inference, the Lieb--Yngvason approach begins directly from the ordering of thermodynamic states under physically admissible transformations.

\subsection{Lieb-Yngvason axioms}

The axiomatic framework of Lieb and Yngvason approaches entropy directly from the phenomenological second law of thermodynamics, without assuming a microscopic model, statistical mechanics, heat, or temperature \cite{Lieb1999physrep}. Its fundamental concept is the relation of \emph{adiabatic accessibility} between equilibrium states. For two states \(X\) and \(Y\), the notation
\begin{equation}
X\prec Y
\end{equation}
means that \(Y\) can be reached from \(X\) by a process whose only net effect on the surroundings is mechanical work. If both \(X\prec Y\) and \(Y\prec X\), the states are adiabatically equivalent, which we denote by
\begin{equation}
X\sim_{\mathrm A}Y.
\end{equation}
The relation \(\prec\) is assumed to satisfy the following axioms.

\paragraph{A1: Reflexivity.}
Every state is adiabatically accessible from itself,
\begin{equation}
X\prec X.
\end{equation}

\paragraph{A2: Transitivity.}
Successive accessible transformations can be composed,
\begin{equation}
X\prec Y,\qquad Y\prec Z
\quad\Longrightarrow\quad
X\prec Z.
\end{equation}

\paragraph{A3: Consistency.}
Accessibility is preserved under composition of systems,
\begin{equation}
X\prec X',
\qquad
Y\prec Y'
\quad\Longrightarrow\quad
(X,Y)\prec(X',Y').
\end{equation}
Here, \((X,Y)\) denotes the compound state formed from the two subsystems.

\paragraph{A4: Scaling invariance.}
If an adiabatic transformation is possible, the corresponding transformation remains possible after scaling the amount of material,
\begin{equation}
X\prec Y
\quad\Longrightarrow\quad
\lambda X\prec\lambda Y,
\qquad
\lambda>0.
\end{equation}

\paragraph{A5: Splitting and recombination.}
A system is adiabatically equivalent to scaled copies containing complementary fractions of the original amount,
\begin{equation}
X
\sim_{\mathrm A}
\left((1-\lambda)X,\lambda X\right),
\qquad
0<\lambda<1.
\end{equation}

\paragraph{A6: Stability.}
An arbitrarily small auxiliary system cannot enable a transformation that would otherwise be impossible. More precisely, if
\begin{equation}
\left(X,\varepsilon Z_0\right)
\prec
\left(Y,\varepsilon Z_1\right)
\end{equation}
for a sequence \(\varepsilon\to0^+\), then
\begin{equation}
X\prec Y.
\end{equation}

These axioms make adiabatic accessibility a preorder compatible with composition and scaling. To represent this ordering by a scalar entropy, one further requires the \emph{comparison hypothesis}: any two states \(X\) and \(Y\) belonging to the same state space are comparable,
\begin{equation}
X\prec Y
\qquad\text{or}\qquad
Y\prec X.
\end{equation}
Under the axioms and comparison hypothesis, there exists an entropy function satisfying the following properties:

\paragraph{E1: Monotonicity.}
The entropy represents the order of adiabatic accessibility,
\begin{equation}
X\prec Y
\quad\Longleftrightarrow\quad
S(X)\leq S(Y).
\end{equation}

\paragraph{E2: Additivity.}
The entropy of a compound system is the sum of the subsystem entropies,
\begin{equation}
S(X,Y)=S(X)+S(Y).
\end{equation}

\paragraph{E3: Extensivity.}
Scaling the amount of material scales the entropy by the same factor,
\begin{equation}
S(\lambda X)=\lambda S(X).
\end{equation}
Together, E1--E3 constitute the entropy principle. Thus, the second law is expressed as an ordering principle: an adiabatic transition is possible precisely when it does not decrease entropy.

The original construction applies to equilibrium states. For general non-equilibrium states, a unique entropy satisfying all of the same properties need not exist. Instead, lower and upper entropies can be defined to delimit which transitions to and from equilibrium states are possible \cite{Lieb2013rspa}. Similarly, systems that cannot be scaled in the ordinary extensive sense can be compared with an auxiliary entropy meter, allowing entropy differences to be assigned without assuming conventional extensivity \cite{Lieb2014rspa}.

\paragraph{Statistical-mechanical interpretation.}
Although statistical mechanics is deliberately absent from the original Lieb--Yngvason derivation, it is instructive to ask when Boltzmann entropy,
\begin{equation}
S_{\mathrm B}(X)=\ln W(X),
\end{equation}
satisfies E1--E3, using the convention \(k_{\mathrm B}=1\) introduced earlier. Monotonicity E1 requires adiabatic accessibility to agree with the ordering of multiplicities,
\begin{equation}
X\prec Y
\quad\Longleftrightarrow\quad
W(X)\leq W(Y).
\end{equation}
Additivity E2 requires multiplicative composition of the accessible microstate spaces,
\begin{equation}
W(X,Y)=W(X)W(Y),
\end{equation}
while extensivity E3 requires
\begin{equation}
W(\lambda X)=W(X)^\lambda.
\end{equation}
Because microscopic multiplicities are integer-valued, this scaling relation is generally interpreted asymptotically in the thermodynamic limit.
Splitting and recombination then imply
\begin{equation}
W(X)
=
W((1-\lambda)X)W(\lambda X).
\end{equation}

These conditions require multiplicative composition and exponential scaling of multiplicity, but they do not uniquely imply multinomial counting. The multinomial coefficient is one particular microscopic realization, corresponding to distinguishable independent constituents assigned among fixed states. Other, non-multinomial multiplicities can satisfy the same asymptotic requirements. Conversely, correlated or non-extensive systems may violate exact factorization or scaling, so the usual Boltzmann entropy may satisfy the Lieb--Yngvason properties only asymptotically, or an extended state description or entropy framework may be required.

\section{Dualities and invariances in generalized entropies}

Much of the study of generalized entropies focuses on the MaxEnt distributions obtained by their constrained maximization. However, observing a particular distribution does not uniquely determine either the entropy functional or the underlying physical mechanism. For example, \(q\)-exponential distributions arise from maximizing Tsallis entropy, and equivalently R\'enyi entropy, under suitable constraints, but they can also be obtained by maximizing Shannon entropy with an appropriate tail constraint \cite{Bercher2008pla}. Tsallis-type power laws have also been derived within standard classical statistical mechanics by assuming separability of the microcanonical distribution in the total system energy \cite{Hanel2005physa}, and from systems coupled to finite heat baths, where the reservoir heat capacity determines the deformation parameter \cite{plastino1994pla,almeida2001physa}. Transformations of this parameter further connect finite-bath and temperature-fluctuation settings with classes of \(q\)-deformed MaxEnt distributions \cite{korbel2017pla}. More broadly, finite reservoir capacity and temperature fluctuations motivate modified canonical weights and entropy formulas that include Boltzmann--Gibbs, R\'enyi, and Tsallis forms as particular cases \cite{Biro2015physa}.

A MaxEnt distribution should therefore be interpreted together with its constraints, composition rules, and physical setting. In this section, we investigate several dualities and invariances through which different entropy formulations or inference settings can lead to the same family of MaxEnt distributions.

\subsection{Calibration invariance of MaxEnt distributions}

The MaxEnt principle has two closely related, but distinct, roles. First, it is an inference procedure used to determine the probability distribution that maximizes a chosen entropy under the available constraints. Second, after the maximizing distribution has been found, the entropy and the corresponding Lagrange multipliers are used to define thermodynamic quantities, such as temperature and thermodynamic potentials. Although the first step determines the most probable distribution, it does not by itself uniquely determine its thermodynamic interpretation \cite{Korbel2021entropy}.

To see this distinction, consider a general entropy \(S(\boldsymbol{p})\), together with the normalization and mean-energy constraints
\begin{equation}
f_0(\boldsymbol{p})
=
\sum_i p_i-1=0,
\qquad
f_E(\boldsymbol{p})
=
\sum_i p_i\epsilon_i-U=0.
\end{equation}
The MaxEnt distribution \(\boldsymbol{p}^\star\) is obtained from the stationary points of
\begin{equation}
\mathcal{L}(\boldsymbol{p})
=
S(\boldsymbol{p})
-\alpha f_0(\boldsymbol{p})
-\beta f_E(\boldsymbol{p}).
\end{equation}
The distribution, therefore, depends on the combination of entropy, constraints, and Lagrange multipliers. Transformations that leave \(\boldsymbol{p}^\star\) unchanged while modifying the corresponding multipliers are called \emph{calibration transformations}. They preserve the inference represented by the MaxEnt distribution, but may change the calibration of the associated thermodynamic quantities.

\paragraph{Monotonic transformations of entropy.}
The first class consists of replacing the entropy by
\begin{equation}
S_c(\boldsymbol{p})
=
c\left(S(\boldsymbol{p})\right),
\end{equation}
where \(c\) is differentiable and satisfies \(c'(x)>0\) on the relevant entropy range. Since \(c\) preserves the ordering of entropy values, \(S\) and \(S_c\) attain their maximum at the same distribution under the same constraints. The stationarity condition for the transformed entropy is
\begin{equation}
c'\left(S(\boldsymbol{p}^\star)\right)
\frac{\partial S}{\partial p_i}
-\alpha_c\frac{\partial f_0}{\partial p_i}
-\beta_c\frac{\partial f_E}{\partial p_i}
=0,
\end{equation}
and comparison with the original condition gives
\begin{equation}
\alpha_c
=
c'\left(S(\boldsymbol{p}^\star)\right)\alpha,
\qquad
\beta_c
=
c'\left(S(\boldsymbol{p}^\star)\right)\beta.
\end{equation}
Thus, the MaxEnt distribution remains invariant, but the multipliers and the thermodynamic quantities identified with them are generally recalibrated.

\noindent\textbf{Tsallis and R\'enyi entropies.}
An important example is the monotonic relation between Tsallis and R\'enyi entropies,
\begin{equation}
R_q(\boldsymbol{p})
=
\frac{1}{1-q}
\ln\left[1+(1-q)S_q(\boldsymbol{p})\right].
\end{equation}
They consequently yield the same MaxEnt distributions when maximized under identical constraints, while their thermodynamic temperatures and potentials need not coincide, which can lead to phenomena such as \emph{self-referential temperature} in Tsallis thermodynamics \cite{Jizba2017pre}. Nevertheless, combinations that transform covariantly, such as the heat \(\text{đ}Q=dS/\beta\), remain invariant under this change of entropy calibration.

\noindent\textbf{Shannon entropy and entropy power.}
A second example is the relation between Shannon entropy and entropy power,
\begin{equation}
\mathcal{P}(\boldsymbol{p})
=
\prod_i\left(\frac{1}{p_i}\right)^{p_i}
=
\exp\left[H(\boldsymbol{p})\right].
\end{equation}
Because the exponential is strictly increasing, maximizing \(\mathcal{P}\) is equivalent to maximizing \(H\). However, their Lagrange multipliers satisfy
\begin{equation}
\alpha_{\mathcal{P}}
=
\mathcal{P}(\boldsymbol{p}^\star)\alpha_H,
\qquad
\beta_{\mathcal{P}}
=
\mathcal{P}(\boldsymbol{p}^\star)\beta_H,
\end{equation}
so even this elementary change of entropy scale modifies the thermodynamic calibration.

\paragraph{Transformations of constraints.}
The second class preserves the entropy but replaces the energetic constraint by a function of the original normalization and energy constraints,
\begin{equation}
f_E(\boldsymbol{p})
\longrightarrow
g\left(f_0(\boldsymbol{p}),f_E(\boldsymbol{p})\right),
\end{equation}
where \(g(0,y)=0\) if and only if \(y=0\). The transformed and original constraints therefore define the same admissible probability distributions. For the transformed Lagrangian
\begin{equation}
\mathcal{L}_g(\boldsymbol{p})
=
S(\boldsymbol{p})
-\alpha_g f_0(\boldsymbol{p})
-\beta_g g\left(f_0(\boldsymbol{p}),f_E(\boldsymbol{p})\right),
\end{equation}
comparison of the stationarity conditions at \(\boldsymbol{p}^\star\) yields
\begin{equation}
\beta
=
\beta_g
\left.\frac{\partial g}{\partial f_E}\right|_{\boldsymbol{p}^\star},
\qquad
\alpha
=
\alpha_g
+
\beta_g
\left.\frac{\partial g}{\partial f_0}\right|_{\boldsymbol{p}^\star}.
\end{equation}
Hence, an equivalent formulation of the constraints leaves the maximizing distribution unchanged but changes the values assigned to its multipliers. This class includes, for example, shifts of the energy spectrum and transformations between alternative representations of the mean energy.

\noindent\textbf{Energy shift.}
For a constant shift of the energy origin,
\begin{equation}
\epsilon_i'
=
\epsilon_i-\epsilon_0,
\qquad
U'
=
U-\epsilon_0,
\end{equation}
the shifted constraint can be written as
\begin{equation}
f_E'(\boldsymbol{p})
=
\sum_i p_i\epsilon_i'-U'
=
f_E(\boldsymbol{p})-\epsilon_0 f_0(\boldsymbol{p}).
\end{equation}
It therefore defines the same admissible distributions whenever normalization holds. The energy multiplier remains unchanged, \(\beta'=\beta\), whereas the normalization multiplier is shifted according to
\begin{equation}
\alpha'
=
\alpha+\beta\epsilon_0,
\end{equation}
corresponding to the shift of the free energy reference level.

\noindent\textbf{Latent escort mean.}
Another example is provided by the escort mean
\begin{equation}
U_q
=
\frac{\sum_i p_i^q\epsilon_i}{\sum_i p_i^q}.
\end{equation}
At \(q=1\), its apparently nonlinear constraint becomes the \emph{latent linear mean}
\begin{equation}
\frac{\sum_i p_i\epsilon_i}{\sum_i p_i}-U=0.
\end{equation}
Multiplication by its denominator gives
\begin{equation}
\sum_i p_i\epsilon_i-U\sum_i p_i
=
f_E(\boldsymbol{p})-U f_0(\boldsymbol{p})=0.
\end{equation}
Thus, once normalization is imposed, the latent mean is equivalent to the ordinary linear mean. Within the MaxEnt procedure, it acts as a shift of the energy spectrum by its average value and changes only the calibration of the normalization multiplier.

Both entropy and constraint calibration transformations can be composed and inverted and therefore possess a group structure \cite{Korbel2021entropy}. They preserve the maximizing distribution while recalibrating the entropy, constraints, and associated multipliers. Thermodynamic quantities must therefore be interpreted relative to the chosen calibration.

\subsection{Hanel--Thurner--Gell-Mann duality between linear and escort averages}

Generalized MaxEnt distributions can be obtained using either ordinary linear constraints or constraints defined through escort probabilities. These two formulations generally employ different entropy functionals and may lead to different thermodynamic interpretations. Nevertheless, for a broad class of generalized distributions, they can be related by a duality that produces the same maximizing probability distribution \cite{hanel2011pnas,hanel2012pnas}.

Consider a trace-form entropy
\begin{equation}
S(\boldsymbol{p})
=
\sum_i s(p_i).
\end{equation}
In the first formulation, the energy is constrained by the ordinary linear mean,
\begin{equation}
U
=
\sum_i p_i\epsilon_i.
\end{equation}
In the second formulation, it is constrained by an escort mean,
\begin{equation}
U^\ast
=
\sum_i P_i\epsilon_i,
\qquad
P_i
=
\frac{u(p_i)}{\sum_j u(p_j)},
\end{equation}
where \(u\) is a positive increasing function. The two formulations correspond to distinct MaxEnt problems because the escort distribution \(\boldsymbol{P}\) assigns different statistical weights to the states than the original distribution \(\boldsymbol{p}\).

Suppose that the generalized canonical distribution is expressed through a generalized logarithm \(\Lambda\), whose inverse is a generalized exponential. The linear-constraint formulation and the escort-constraint formulation reproduce the same MaxEnt distribution when their generalized logarithms are related by
\begin{equation}
\Lambda^\ast(x)
=
T_\nu\left(\Lambda(x)\right),
\qquad
T_\nu(x)
=
\frac{x}{1+\nu x}.
\end{equation}
Where the generalized logarithms are nonzero, this can equivalently be written as
\begin{equation}
\frac{1}{\Lambda^\ast(x)}
=
\frac{1}{\Lambda(x)}+\nu.
\end{equation}
The logarithms \(\Lambda\) and \(\Lambda^\ast\) generate dual entropy representations: one is maximized under the linear energy constraint, while the other is maximized under the corresponding escort constraint. For a given entropy, the duality relation specifies how the logarithms in the two formulations must be related.

Consistency between the two MaxEnt procedures also fixes the escort distribution. It cannot be chosen independently of the entropy but must have the form
\begin{equation}
P_i
=
\frac{p_i+\nu s(p_i)}
{1+\nu S(\boldsymbol{p})}.
\end{equation}
Indeed, the denominator follows from normalization,
\begin{equation}
\sum_i\left[p_i+\nu s(p_i)\right]
=
1+\nu S(\boldsymbol{p}).
\end{equation}
The escort probabilities are therefore determined by the entropy density \(s\) and the duality parameter \(\nu\). Admissible choices must satisfy \(p_i+\nu s(p_i)\geq0\) for every state and \(1+\nu S(\boldsymbol{p})>0\), so that \(\boldsymbol{P}\) is a valid probability distribution. For \(\nu=0\), one recovers \(P_i=p_i\), and the distinction between the two constraint prescriptions disappears.

\noindent\textbf{Tsallis duality.}
For the \(q\)-logarithm
\begin{equation}
\ln_q(x)
=
\frac{x^{1-q}-1}{1-q},
\end{equation}
choosing \(\nu=1-q\) gives
\begin{equation}
T_{1-q}\left(\ln_q x\right)
=
\ln_{2-q}x.
\end{equation}
Thus, the general logarithmic duality reduces to the familiar additive duality
\begin{equation}
q^\ast=2-q.
\end{equation}
Maximizing the corresponding dual entropy with escort constraints yields the same \(q\)-exponential family as the linear-constraint representation, although the entropy functional and constraint prescription are different.

\noindent\textbf{Escort entropies.}
Standard R\'enyi or Tsallis entropies and divergences can also be evaluated on escort distributions. This produces two-parameter families of escort entropies and associated canonical distributions \cite{Bercher2011escort}. In these families, transforming the probabilities to their escort form can be accompanied by a corresponding transformation of the entropy representation.

The duality establishes equivalent linear- and escort-constraint representations of the same generalized canonical distribution. The appropriate representation must be selected from the physical meaning of the measured averages and the thermodynamic relations of the system.

\subsection{Thermodynamics of exponential KN averages}

The arithmetic mean is not the only consistent way to average a set of values. Kolmogorov and Nagumo independently introduced the broad class of quasi-arithmetic means \cite{kolmogorov1930moyenne,Nagumo1930jjm},
\begin{equation}
\langle A\rangle_\phi
=
\phi^{-1}\left(\sum_i p_i\phi(A_i)\right),
\end{equation}
where \(\phi\) is continuous and strictly monotonic. The ordinary linear mean follows from \(\phi(x)=x\). Kolmogorov--Nagumo (KN) averages have subsequently provided a useful framework for generalized thermostatistics \cite{Czachor2002pla}. In particular, the Legendre-transform structure and the corresponding thermodynamic relations can remain consistent when linear averages are replaced by suitable KN averages \cite{Scarfone2016pla}.

\paragraph{Exponential KN averages.}
The exponential family is generated by
\begin{equation}
\phi_\gamma(x)
=
\frac{\exp(\gamma x)-1}{\gamma},
\end{equation}
and gives the average
\begin{equation}
\langle A\rangle_\gamma
=
\frac{1}{\gamma}
\ln\left(\sum_i p_i\exp(\gamma A_i)\right).
\end{equation}
The arithmetic mean is recovered in the limit \(\gamma\to0\). Moreover, since the exponential KN average is proportional to the cumulant-generating function, it incorporates all cumulants of \(A\), weighted by increasing powers of \(\gamma\).

To preserve the physical units of energy, the corresponding internal energy is defined as
\begin{equation}
U_\gamma^\beta(\boldsymbol{p})
=
\frac{1}{\beta\gamma}
\ln\left(\sum_i p_i\exp(\gamma\beta\epsilon_i)\right),
\end{equation}
where \(\beta\) has units of inverse energy. The natural entropy associated with this average is a rescaled R\'enyi entropy,
\begin{equation}
\mathcal{R}_\gamma(\boldsymbol{p})
=
\frac{1}{\gamma(1-\gamma)}
\ln\left(\sum_i p_i^{1-\gamma}\right)
=
\frac{R_{1-\gamma}(\boldsymbol{p})}{1-\gamma}.
\end{equation}
The final equality relates the convention used here to the standard R\'enyi entropy defined earlier. For the MaxEnt construction below, we assume \(\gamma<1\), with \(\gamma=0\) understood as a limit, so this positive rescaling leaves the maximizing distribution unchanged while simplifying the resulting thermodynamic relations. This normalization also connects continuously, up to a distribution-independent constant, to the normalized Burg entropy in the limit \(\gamma\to1\).

\paragraph{MaxEnt distribution.}
Consider maximizing \(\mathcal{R}_\gamma(\boldsymbol{p})\) under normalization and a fixed exponential KN energy \(U_\gamma^\beta\). The corresponding Lagrangian is
\begin{equation}
\mathcal{L}(\boldsymbol{p})
=
\mathcal{R}_\gamma(\boldsymbol{p})
-\alpha\left(\sum_i p_i-1\right)
-\lambda\left[U_\gamma^\beta(\boldsymbol{p})-\overline{U}_\gamma^\beta\right].
\end{equation}
Although both the entropy and energy constraint are generalized, the resulting equilibrium distribution is the ordinary Boltzmann distribution \cite{Morales2023njp},
\begin{equation}
p_i^\star
=
\frac{\exp(-\beta\epsilon_i)}{Z(\beta)},
\qquad
Z(\beta)
=
\sum_i\exp(-\beta\epsilon_i).
\end{equation}
Thus, the generalized entropy and generalized averaging prescription compensate for one another. The equilibrium statistics remain Boltzmannian, while the thermodynamic entropy, internal energy, and potentials differ from their conventional forms.

\paragraph{Equilibrium thermodynamics.}
For the Boltzmann distribution, the generalized free energy becomes
\begin{eqnarray}
\mathcal{F}_\gamma^\beta
&=&
U_\gamma^\beta(\boldsymbol{p}^\star)
-\frac{1}{\beta}\mathcal{R}_\gamma(\boldsymbol{p}^\star)
\nonumber\\
&=&
\frac{1}{(\gamma-1)\beta}
\ln\left[
\sum_i\exp\left((\gamma-1)\beta\epsilon_i\right)
\right].
\end{eqnarray}
Introducing
\begin{equation}
\beta'=(1-\gamma)\beta,
\end{equation}
this can be written as
\begin{equation}
\mathcal{F}_\gamma^\beta
=
F^{\beta'},
\end{equation}
where \(F^{\beta'}=-\ln Z(\beta')/\beta'\) is the ordinary Helmholtz free energy evaluated at inverse temperature \(\beta'\). The KN formalism therefore effectively relates thermodynamic potentials at two different temperatures.

The equilibrium R\'enyi entropy makes this relation explicit:
\begin{eqnarray}
\mathcal{R}_\gamma(\boldsymbol{p}^\star)
&=&
\frac{1}{\gamma(1-\gamma)}
\left[
\ln Z(\beta')
-(1-\gamma)\ln Z(\beta)
\right]
\nonumber\\
&=&
-\frac{\beta}{\gamma}
\left(
F^{\beta'}-F^\beta
\right).
\end{eqnarray}
Equivalently,
\begin{equation}
\mathcal{R}_\gamma(\boldsymbol{p}^\star)
=
\beta^2
\frac{F^{\beta'}-F^\beta}{\beta'-\beta}.
\end{equation}
This difference quotient approaches the usual thermodynamic relation between Shannon entropy and free energy as \(\gamma\to0\), for which \(\beta'\to\beta\).

\paragraph{Temperature-quench interpretation.}
The relation \(\beta'=(1-\gamma)\beta\) admits a direct equilibrium interpretation. Consider a system initially equilibrated with a bath at inverse temperature \(\beta\), which is suddenly quenched to a bath at inverse temperature \(\beta'\). The R\'enyi entropy of the initial equilibrium state is related to the maximum work that can be extracted as the system relaxes to the new equilibrium, with the corresponding temperature change providing the appropriate normalization \cite{Baez2022entropy,Morales2023njp}. The parameter \(\gamma\), therefore, quantifies the relative change of bath temperature rather than a deformation of the equilibrium probability distribution.

Exponential KN averages also connect this thermodynamic framework with multifractal scaling and admit non-equilibrium extensions satisfying generalized forms of the second law and the \(H\)-theorem \cite{Morales2023njp}. These extensions lie beyond the equilibrium focus here. At equilibrium, generalized averaging and R\'enyi thermodynamics can preserve the ordinary Boltzmann distribution.

\subsection{Generalized entropies for nonlinear dynamics}

The preceding subsections varied the entropy or constraints while preserving the stationary distribution. A complementary perspective begins with the dynamics: changing its probability dependence can modify the stationary distribution and select the entropy that provides its thermodynamically consistent description. Nonlinear master and Fokker--Planck equations have been studied in connection with anomalous diffusion, generalized thermostatistics, biological coordination, and nonextensive Markov processes \cite{Frank2000physa,Curado2003pre,Frank2004jbp,Peng2020pre}. Here, we focus on the relation between nonlinear Markovian dynamics and generalized entropy developed in Ref.~\cite{Korbel2021njp}.

\paragraph{Nonlinear master equation.}
Consider a finite set of states labeled by \(m,n,\ldots\), with probability distribution \(\boldsymbol{p}\) and state energies \(\epsilon_m\). The average internal energy retains its conventional linear form,
\begin{equation}
U
=
\sum_m p_m\epsilon_m.
\end{equation}
The dynamics is described by the nonlinear master equation
\begin{equation}
\dot p_m
=
\sum_n
\left[
w_{mn}\Omega(p_n)
-
w_{nm}\Omega(p_m)
\right],
\end{equation}
where \(w_{mn}\) are generalized transition rates and \(\Omega(p)\) determines how the probability of the departure state affects the transition current. The antisymmetric form of the probability currents guarantees normalization,
\begin{equation}
\sum_m\dot p_m=0.
\end{equation}
For \(\Omega(p)=p\), the usual linear master equation is recovered.

The information entropy is taken from the sum-class,
\begin{equation}
S(\boldsymbol{p})
=
f\left(\sum_m g(p_m)\right),
\end{equation}
where \(f\) is strictly increasing and \(g\) is concave. Trace-form entropies correspond to the particular choice \(f(x)=x\). Maximizing this entropy subject to normalization and a fixed mean energy gives the equilibrium distribution
\begin{equation}
p_m^\star
=
\left(g'\right)^{-1}
\left[
\frac{\alpha+\beta\epsilon_m}
{C_f(\boldsymbol{p}^\star)}
\right],
\end{equation}
where
\begin{equation}
C_f(\boldsymbol{p})
=
f'\left(\sum_m g(p_m)\right).
\end{equation}

\paragraph{Local detailed balance.}
Suppose that the system is coupled to reservoirs labeled by \(\ell\), with inverse temperatures \(\beta^\ell=1/T^\ell\), and that
\begin{equation}
w_{mn}
=
\sum_\ell w_{mn}^{\ell}.
\end{equation}
For each reservoir, the local equilibrium distribution \(\boldsymbol{p}^{\star,\ell}\) is defined as the distribution maximizing \(S(\boldsymbol{p})\) at the corresponding mean energy. Generalized local detailed balance requires each reservoir-induced probability current to vanish at its local equilibrium,
\begin{equation}
w_{mn}^{\ell}\Omega\left(p_n^{\star,\ell}\right)
=
w_{nm}^{\ell}\Omega\left(p_m^{\star,\ell}\right).
\end{equation}
Once the nonlinear dynamics and entropy are related consistently, this condition takes the conventional energetic form
\begin{equation}
\frac{w_{mn}^{\ell}}{w_{nm}^{\ell}}
=
\exp\left[
-\beta^\ell\left(\epsilon_m-\epsilon_n\right)
\right].
\end{equation}
Here, \(\beta^\ell\) is the physically calibrated inverse temperature. In the MaxEnt stationarity condition, the corresponding energy multiplier is \(C_f(\boldsymbol{p}^{\star,\ell})\beta^\ell\): the normalization multiplier cancels in the detailed-balance ratio, while the common factor \(C_f\) is absorbed into the calibration of \(\beta^\ell\).

\paragraph{Entropy selected by the dynamics.}
The central result is obtained by requiring simultaneously nonlinear Markovian evolution, local detailed balance, and a non-negative entropy production. These requirements fix the relation between the nonlinear kinetic function and the entropy density \cite{Korbel2021njp},
\begin{equation}
\Omega(p)
=
\exp\left[-g'(p)\right].
\end{equation}
Thus, the entropy and nonlinear dynamics cannot be selected independently. Given \(\Omega\), the thermodynamically consistent sum-class entropy is
\begin{equation}
S(\boldsymbol{p})
=
f\left[
-\sum_m
\int_0^{p_m}
\ln\Omega(z)\,dz
\right],
\end{equation}
up to constants and affine terms that do not affect the MaxEnt distribution.

Conversely, substituting the relation \(\Omega(p)=\exp[-g'(p)]\) into the nonlinear master equation gives
\begin{equation}
\dot p_m
=
\sum_{n,\ell}
\left[
w_{mn}^{\ell}\exp\left(-g'(p_n)\right)
-
w_{nm}^{\ell}\exp\left(-g'(p_m)\right)
\right].
\end{equation}
The form of the generalized entropy therefore determines the nonlinear probability dependence of the transition currents.

\paragraph{Stationary distribution and MaxEnt.}
For a system coupled to a single reservoir, detailed balance at stationarity implies
\begin{equation}
\frac{\Omega(p_m^\star)}
{\Omega(p_n^\star)}
=
\exp\left[
-\beta\left(\epsilon_m-\epsilon_n\right)
\right].
\end{equation}
Consequently,
\begin{equation}
\Omega(p_m^\star)
=
\exp\left(-\alpha-\beta\epsilon_m\right),
\end{equation}
where \(\alpha\) enforces normalization. Using \(\Omega(p)=\exp[-g'(p)]\), this stationary condition is equivalent to
\begin{equation}
g'(p_m^\star)
=
\alpha+\beta\epsilon_m,
\end{equation}
up to an overall calibration of the multipliers by \(C_f(\boldsymbol{p}^\star)\). This is precisely the stationarity condition obtained by maximizing the associated entropy under normalization and fixed mean energy. The stationary distribution of the nonlinear dynamics is therefore the MaxEnt distribution of the entropy selected by that dynamics.

\noindent\textbf{Linear dynamics and Shannon entropy.}
For the ordinary linear master equation, \(\Omega(p)=p\), the consistency relation gives
\begin{equation}
g'(p)
=
-\ln p.
\end{equation}
Integration yields
\begin{equation}
g(p)
=
-p\ln p+p,
\end{equation}
where the linear term does not affect the entropy under normalization. Choosing \(f(x)=x\) consequently recovers Shannon entropy and the stationary Boltzmann distribution.

\noindent\textbf{Other nonlinear examples.}
Different choices of the entropy density generate different nonlinear transition factors. For example, the Fermi--Dirac entropy density
\begin{equation}
g_{\mathrm{FD}}(p)
=
-p\ln p-(1-p)\ln(1-p)
\end{equation}
gives
\begin{equation}
\Omega_{\mathrm{FD}}(p)
=
\frac{p}{1-p}.
\end{equation}
The divergence as \(p\to1\) reflects the exclusion constraint. Conversely, the Bose--Einstein entropy density
\begin{equation}
g_{\mathrm{BE}}(p)
=
(1+p)\ln(1+p)-p\ln p
\end{equation}
leads to
\begin{equation}
\Omega_{\mathrm{BE}}(p)
=
\frac{p}{1+p},
\end{equation}
which represents an inclusion effect. Substitution into the stationary condition \(\Omega(p_m^\star)=\exp(-\alpha-\beta\epsilon_m)\) recovers the corresponding Fermi--Dirac and Bose--Einstein stationary distributions.

As another example, Burg entropy has density \(g_{\mathrm{BU}}(p)=\ln p\), and therefore corresponds to
\begin{equation}
\Omega_{\mathrm{BU}}(p)
=
\exp\left(-\frac{1}{p}\right).
\end{equation}
Its stationary distribution satisfies
\begin{equation}
p_m^\star
=
\frac{1}{\alpha+\beta\epsilon_m}.
\end{equation}
More generally, power-law entropy densities generate nonlinear master equations relevant to chemical reaction networks, anomalous transport, and other complex systems \cite{Korbel2021njp}. Generalized entropies can therefore arise from the probability dependence of the dynamics, as well as from microscopic multiplicities, axiomatic requirements, or alternative constraints. The transition rates select the entropy thermodynamically compatible with the stationary distribution.

\section{Concluding remarks}

In this chapter, we examined entropy from several complementary viewpoints. Starting from Boltzmann's formula, we showed how entropy follows from the multiplicity of coarse-grained states. Multinomial counting leads to Shannon entropy, while different particle statistics, emergent structures, and history-dependent processes lead to different entropy forms and MaxEnt distributions.

We then considered axiomatic approaches, which ask what properties entropy should satisfy without first specifying a microscopic model. These approaches clarify the roles of composition, phase-space growth, statistical inference, and thermodynamic accessibility. Finally, we discussed several cases in which different entropies, constraints, or dynamics lead to the same stationary distribution.

These observations suggest that generalized entropy should not be chosen solely from the shape of an observed distribution. The choice of entropy should, whenever possible, be supported by the system's microscopic counting, dynamics, constraints, or thermodynamic properties. Many open questions remain, particularly concerning how these different viewpoints are related and how they can be applied in empirical studies of complex systems.

\section*{Acknowledgements}

This chapter grew out of lectures presented at the \emph{School on Information, Noise, and Physics of Life 2022} and the \emph{Ising Lectures 2026}. I thank the organizers of both events, particularly Velimir Ili\'{c} and Yurij Holovatch, for their support.

This work was supported by the Austrian Science Fund (FWF) through Grant No.~10.55776/P34994 and EFP5 ReMASS, and by the Austrian Federal Ministry for Climate Action, Environment, Energy, Mobility, Innovation, and Technology under GZ~2023-0.841.266, through the Postdoc Program for Complexity Science and Data Competence.

\bibliographystyle{apsrev4-2-titles}
\bibliography{bibliography_arxiv}

\end{document}